\documentclass[useAMS,usenatbib,usegraphicx]{mn2e}
\usepackage{amsmath,amsfonts,amssymb}

\newcommand{\lya}{Ly$\alpha$}
\newcommand{\1}{$^{-1}$}

\newcommand{\3}{$^{-3}$}
\newcommand{\hm}{$h^{-1}$}
\newcommand{\kms}{km s\1}
\newcommand{\msun}{M$_{\sun}$}

\newcommand{\apj}{ApJ}
\newcommand{\apjl}{ApJL}
\newcommand{\apjs}{ApJS}
\newcommand{\mnras}{MNRAS}
\newcommand{\pasj}{PASJ}
\newcommand{\pasp}{PASP}
\newcommand{\sci}{Science}
\newcommand{\ssr}{SSR}

\newcommand{\ion}[2]{#1\,{\small{#2}}}
\newcommand{\hi}{\ion{H}{I}}
\newcommand{\hei}{\ion{He}{I}}
\newcommand{\heii}{\ion{He}{II}}

\newcommand{\cii}{\ion{C}{II}}
\newcommand{\ciii}{\ion{C}{III}}
\newcommand{\civ}{\ion{C}{IV}}

\newcommand{\niii}{\ion{N}{III}}
\newcommand{\niv}{\ion{N}{IV}}
\newcommand{\nv}{\ion{N}{V}}

\newcommand{\oi}{\ion{O}{I}}
\newcommand{\oii}{\ion{O}{II}}
\newcommand{\oiii}{\ion{O}{III}}
\newcommand{\oiv}{\ion{O}{IV}}
\newcommand{\ov}{\ion{O}{V}}
\newcommand{\ovi}{\ion{O}{VI}}
\newcommand{\ovii}{\ion{O}{VII}}
\newcommand{\oviii}{\ion{O}{VIII}}
\newcommand{\nei}{\ion{Ne}{I}}

\newcommand{\neiv}{\ion{Ne}{IV}}
\newcommand{\nev}{\ion{Ne}{V}}
\newcommand{\nevi}{\ion{Ne}{VI}}
\newcommand{\nevii}{\ion{Ne}{VII}}
\newcommand{\neviii}{\ion{Ne}{VIII}}

\newcommand{\mgii}{\ion{Mg}{II}}
\newcommand{\mgiii}{\ion{Mg}{III}}
\newcommand{\mgvi}{\ion{Mg}{VI}}
\newcommand{\mgvii}{\ion{Mg}{VII}}
\newcommand{\mgviii}{\ion{Mg}{VIII}}
\newcommand{\mgix}{\ion{Mg}{IX}}
\newcommand{\sili}{\ion{Si}{I}}
\newcommand{\silii}{\ion{Si}{II}}
\newcommand{\siliii}{\ion{Si}{III}}
\newcommand{\siliv}{\ion{Si}{IV}}
\newcommand{\silv}{\ion{Si}{V}}
\newcommand{\silvi}{\ion{Si}{VI}}
\newcommand{\silvii}{\ion{Si}{VII}}
\newcommand{\silviii}{\ion{Si}{VIII}}
\newcommand{\silix}{\ion{Si}{IX}}
\newcommand{\silx}{\ion{Si}{X}}
\newcommand{\silxi}{\ion{Si}{XI}}
\newcommand{\silxii}{\ion{Si}{XII}}
\newcommand{\si}{\ion{S}{I}}
\newcommand{\siii}{\ion{S}{III}}
\newcommand{\sv}{\ion{S}{V}}
\newcommand{\six}{\ion{S}{IX}}
\newcommand{\sx}{\ion{S}{X}}
\newcommand{\sxi}{\ion{S}{XI}}
\newcommand{\sxii}{\ion{S}{XII}}
\newcommand{\sxiii}{\ion{S}{XIII}}

\newcommand{\caii}{\ion{Ca}{II}}
\newcommand{\cavi}{\ion{Ca}{VI}}
\newcommand{\cavii}{\ion{Ca}{VII}}
\newcommand{\fei}{\ion{Fe}{I}}
\newcommand{\feii}{\ion{Fe}{II}}

\newcommand{\fevii}{\ion{Fe}{VII}}
\newcommand{\feviii}{\ion{Fe}{VIII}}
\newcommand{\feix}{\ion{Fe}{IX}}
\newcommand{\fex}{\ion{Fe}{X}}
\newcommand{\fexi}{\ion{Fe}{XI}}
\newcommand{\fexii}{\ion{Fe}{XII}}
\newcommand{\fexiii}{\ion{Fe}{XIII}}
\newcommand{\fexiv}{\ion{Fe}{XIV}}
\newcommand{\fexv}{\ion{Fe}{XV}}
\newcommand{\fexvi}{\ion{Fe}{XVI}}
\newcommand{\fexxv}{\ion{Fe}{XXV}}
\newcommand{\fexx}{\ion{Fe}{XX}}
\newcommand{\fexxvi}{\ion{Fe}{XXVI}}

\newcommand{\owls}{OWLS}
\newcommand{\default}{\emph{REF}}         
\newcommand{\nosn}{\emph{NOSN}}           
\newcommand{\nosnz}{\emph{NOSN\_NOZCOOL}} 
\newcommand{\zcool}{\emph{NOZCOOL}}       
\newcommand{\wmom}{\emph{WVCIRC}}         
\newcommand{\agn}{\emph{AGN}}
\newcommand{\noagb}{\emph{NOAGB\_NOSNIa}}

\newcommand{\cloudy}{{\sc CLOUDY}}
\newcommand{\xray}{X-ray}

\title[Diffuse gas cooling]{How the diffuse Universe cools}
\author[Bertone et al.]{Serena Bertone$^{1,2}$\thanks{E-mail: serena.bertone@aviospace.com}, Anthony Aguirre$^{2}$ and Joop Schaye$^{3}$ \\
$^{1}$ Aviospace S.r.l., via Pier Carlo Boggio 59/61, 10138 Torino, Italy \\
$^{2}$ Santa Cruz Institute for Particle Physics, University of California, 1156 High Street, Santa Cruz CA 95064, USA \\
$^{3}$ Leiden Observatory, Leiden University, P.O. Box 9513, 2300 RA Leiden, The Netherlands }

\begin{document}

\voffset=-0.8in

\date{Accepted by MNRAS}
\pagerange{\pageref{firstpage}--\pageref{lastpage}} \pubyear{2013}
\maketitle
\label{firstpage}

\begin{abstract}
In this work we investigate the cooling channels of diffuse gas (i.e. $n_{\rm H} < 0.1$ cm\3) in cosmology. We aim to identify the wavelengths where most of the energy is radiated in the form of emission lines or continuum radiation, and the main elements and ions responsible for the emission.
We use a subset of cosmological, hydrodynamical runs from the \owls\ project to calculate the emission of diffuse gas and its evolution with time.
We find that at $z=0$ ($z=2$) about 70 (80) per cent of the energy emitted by diffuse gas is carried by emission lines, with the continuum radiation contributing the remainder.
Hydrogen lines in the Lyman series are the primary contributors to the line emission, with a share of 16 (20) per cent. Oxygen lines are the main metal contributors at high redshift, while silicon, carbon and iron lines are strongest at low redshift, when the contributions of AGB stars and supernova Ia explosions to the metal budget become important and when there is more hot gas.
The ionic species carrying the most energy are \oiii, \cii, \ciii, \silii, \siliii, \feii\ and \siii.
The great majority of energy is emitted in the UV band ($\lambda =100-4000$ \AA), both as continuum radiation and line emission. With almost no exception, all the strongest lines fall in this band.  At high energies, continuum radiation is dominant (e.g., 80 per cent in the \xray\ band), while lines contribute progressively more at lower energies.
While the results do depend on the details of the numerical implementation of the physical processes modeled in the simulations, the comparison of results from different simulations demonstrates that the variations are overall small, and that the conclusions are fairly robust.
Given the overwhelming importance of UV emission for the cooling of diffuse gas, it is desirable to build instruments dedicated to the detection and characterisation of diffuse UV emission.

\end{abstract}

\renewcommand{\ion}[2]{#1\,{\scriptsize{#2}}}

\begin{keywords}
intergalactic medium -- diffuse radiation -- radiation mechanisms: thermal -- method: numerical
\end{keywords}

\section{Introduction}
\label{intro}

The evolution of the Universe is determined by the complex interplay of a large number of processes that involve the exchange of mass and energy. A global view of these mass and energy flows has been taken to investigate the universal content of baryonic mass \citep{persic1992, fukugita1998, fukugita2004}, energy \citep{fukugita2004}, metals \citep{pettini2004,bouche2007} and star formation \citep{madau1996}. These quantities give us useful insights in to the evolution of the physical Universe as a whole and they help us understand how individual processes contribute to the evolution.

In this work, we aim to investigate a key part of this global energy interchange: radiative cooling of diffuse gas on cosmological scales.  In particular, we aim to identify the cooling channels that dominate the energy budget of the diffuse gas that traces the large-scale structure of the Universe and that feeds galaxies and fuels star formation through accretion processes. For our purposes, we define this diffuse matter as the sum of three main components: i) the intergalactic medium (IGM), defined as the low-density gas that is not gravitationally bound to galaxies and their haloes; ii) the circum-galactic medium (CGM), which is denser than the IGM, resides in the immediate surroundings of galaxies, and is bound to galaxy haloes; and iii) the intra-cluster medium (ICM), which is the hot, shock-heated gas that accumulates in the gravitational potentials of groups and clusters. Both the CGM and the ICM are gravitationally bound to haloes and represent the densest fraction of diffuse gas.
While the distinctions between these components are often not clear in observations, they can be more cleanly defined in simulations. In the rest of the paper, we will consider gas in simulations as ``diffuse" if its hydrogen number density is $n_{\rm H}<0.1$~cm\3.

Thermal and kinetic energy are injected in to the diffuse gas by gravity, by galactic winds powered by stellar winds and supernova explosions, and by feedback due to black hole accretion (AGN feedback), cosmic rays and other processes. The gas cools via a variety of mechanisms and emits radiation in a broad range of wavelengths. The cooling radiation can in turn heat up gas locally or far away, can be absorbed and re-emitted at lower frequencies by dust, or (if the mean free path of photons is very large) can contribute to the background radiation field.

Cooling processes within galaxies are thought to contribute substantially to the diffuse infrared, optical, UV and \xray\ backgrounds and are an essential ingredient in galaxy formation. These are still not well understood, but can be somewhat constrained and quantified by observation.
The radiation produced by the cooling of diffuse gas, on the other hand, is very difficult to observe directly because of its very low surface brightness \citep[e.g.][]{furlanetto2003, furlanetto2004, furlanetto2005, yoshikawa2003, fang2005, bertone2010a, bertone2010b, bertone2011, goerdt2010, faucher2010, vandevoort2012, rosdahl2012}. Currently, the IGM can therefore be more effectively studied in absorption rather than in emission.

In this work, we aim to provide a global view of how cooling occurs in diffuse gas. To this end, we use a subset of runs from the OverWhelmingly Large Simulations (\owls\ hereafter) project \citep{schaye2010} to address a number of questions concerning the emission and cooling processes of diffuse gas, including:
i) Which transitions, ions, and gas phases are responsible for most of the cooling radiation, and should thus be the focus of theoretical modelling efforts?
ii) Which wavebands and emission lines carry most of the energy emitted by diffuse gas through cooling and fluorescence radiation, and should thus be primary targets for observational studies?
iii) What is the relative importance of continuum emission and line emission?
iv) What is the global emission of diffuse gas, and how does it change with time? How much energy is stored in the IGM?
Since our focus is diffuse gas, we will ignore emission by dust and by gas with densities exceeding $n_{\rm H} = 0.1$~cm\3.

The paper is organized as follows.
Section~\ref{tables} describes the emissivity tables used to calculate the gas emission and shows the dependence of the emissivity on temperature for a large sample of ions, individual emission lines and the continuum.
Section~\ref{owlproj} discusses the simulations and the methodology used to calculate the gas emission.
Section~\ref{volume} presents results for the fraction of energy emitted by different ions, elements and the continuum in the reference simulation set (\default\ in the following) at $z=0$ and $z=2$.
In Section \ref{sims} we compare results from different simulations to understand the effect of varying the physical prescriptions. Finally, we draw our conclusions in Section~\ref{conclusion} and present a series of numerical convergence tests in the Appendix.

\section{The emissivity tables}
\label{tables}

\begin{table}
\centering
\caption{Adopted solar abundances, from \citet{allende2001}, \citet{allende2002} and \citet{holweger2001}.}
\begin{tabular}{l c | l c}
\hline
Element & $n_i / n_{\rm H}$ & Element & $n_i / n_{\rm H}$\\
\hline
H & 1          & Mg & 3.47$\times 10^{-5}$ \\
He & 0.1         & Si & 3.47$\times 10^{-5}$ \\
C & 2.46$\times 10^{-4}$ & S & 1.86$\times 10^{-5}$ \\
N & 8.51$\times 10^{-5}$ & Ca & 2.29$\times 10^{-6}$ \\
O & 4.90$\times 10^{-4}$ & Fe & 2.82$\times 10^{-5}$ \\
Ne & 1.00$\times 10^{-4}$ &  &            \\
\hline
\end{tabular}
\label{table_abund}
\end{table}

In this Section we describe the emissivity tables used to calculate the contribution of different species to the total emission. In particular, in Section \ref{ions} we show the temperature dependence of the emissivity for all ions, and for a sample of the strongest individual emission lines for each ion; Section \ref{cont_em} describes the emissivity tables for the continuum.

The tables have been created using \cloudy\footnote{Version c07.02.02, last described in \citealt{ferland1998} and released in July 2008. See \texttt{http://www.nublado.org/}.} as a function of gas temperature (in bins of $\Delta {\rm Log}_{10}T=0.05$ for $10^2\,$K $< T < 10^{8.5}\,$K), hydrogen number density (in bins of $\Delta {\rm Log}_{10} n_{\rm H}=0.2$ for $10^{-8}\,$cm\3\ $< n_{\rm H} < 10$~cm\3, but note that we will ignore emission from gas with $n_{\rm H} > 10^{-1}\,$cm\3), and redshift.

The adopted solar abundance is $Z_{\sun} =0.0127$, corresponding to the value obtained using the default abundance set of \cloudy\ (see Table \ref{table_abund}). This set combines the abundances of \citet{allende2001}, \citet{allende2002} and \citet{holweger2001} and can differ significantly from the estimates of \citet{lodders2003}.
In particular, the abundance of oxygen given by \citet{lodders2003} is about 20 per cent higher than in \cloudy.
Note that variations in the adopted solar abundances may introduce variations in the computed emissivity per unit gas mass, but not for the simulation results, which are instead calculated using the absolute abundances predicted by the simulations.

The tables created with \cloudy\ assume the gas to be dust-free, optically thin and in (photo-)ionisation equilibrium in the presence of a uniform, evolving cosmic UV/\xray\ background radiation contributed by both galaxies and quasars \citep{haardt2001} and the cosmic microwave background spectrum. While dust emission is an extremely important cooling channel for the interstellar medium, its contribution to the cooling of the diffuse IGM is probably much smaller, particularly at higher temperatures where the dust is efficiently destroyed.
On average, local sources of ionising radiation are expected to become more important than the background for gas clouds with neutral hydrogen column densities $> 10^{17}\,{\rm cm}^{-2}$ \citep{schaye2006}, which are expected to have volume densities slightly below $n_{\rm H} = 10^{-1}\,{\rm cm}^{-3}$ \citep{schaye2001}, the highest gas density we consider. As ionising radiation suppresses cooling rates due to the removal of bound electrons \citep{efstathiou1992, wiersma2009a}, we may have slightly overestimated the cooling rates for high-density gas, particularly for temperatures $\la 10^5\,$K \cite[see][]{wiersma2009a}. Ionisation equilibrium is expected to hold for photo-ionised regions and for dense gas in the centres of clusters (see \citealt{bertone2008} for a review).
The gas is prevalently photo-ionised at low density and low temperatures and is collisionally ionised at high densities and temperatures.
The equilibrium assumption can break down for shock-heated gas. This might happen, for example, for warm-hot gas in the outskirts of groups and clusters, as argued by  \citet{yoshida2005,yoshikawa2006,cenfang2006,gnat2007, gnat2009}. However, since these studies ignored photo-ionisation, they may have overestimated the importance of non-equilibrium ionisation.

\subsection{Line emissivity}
\label{ions}

\begin{figure*}
\centering
\includegraphics[width=0.9\textwidth]{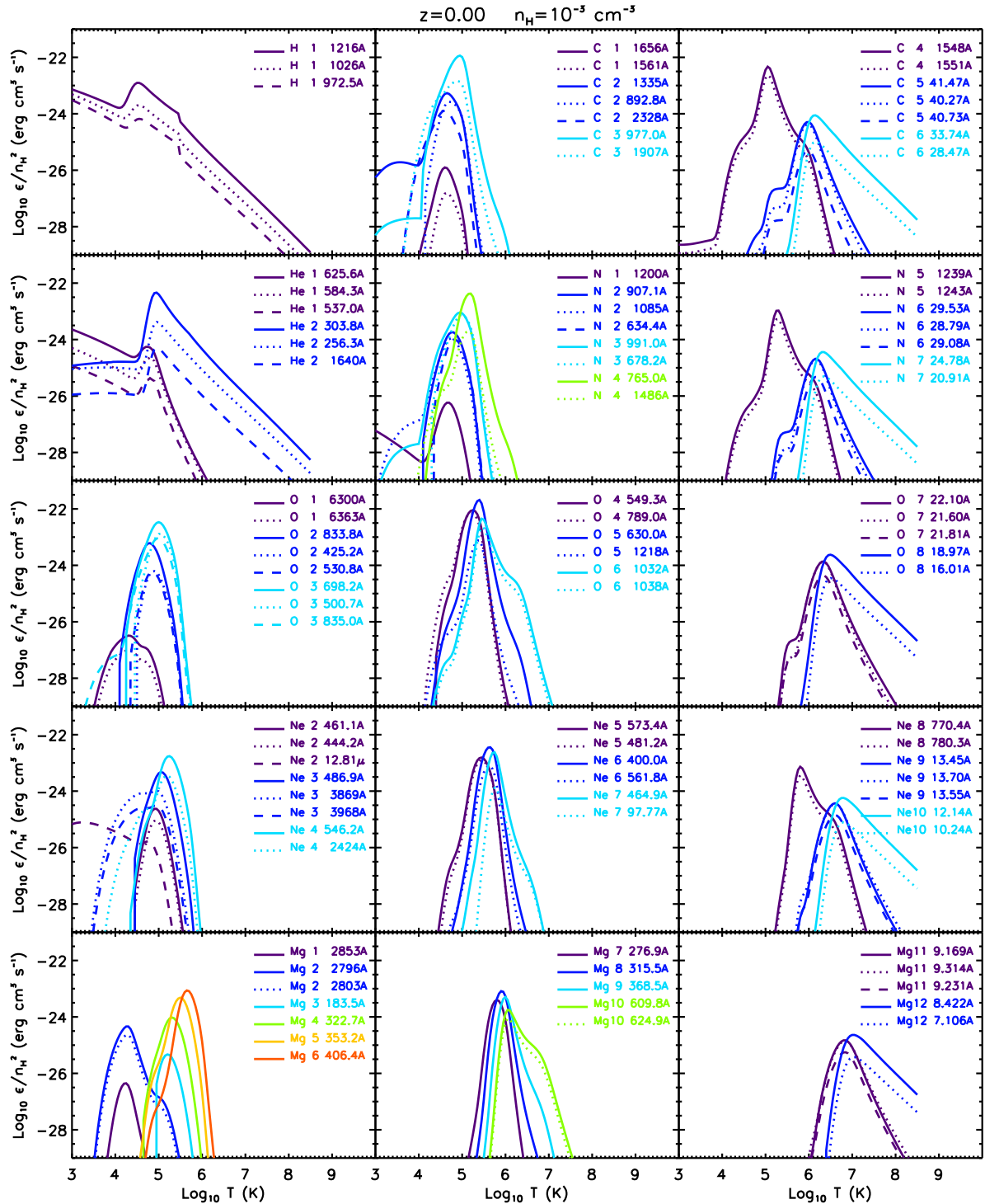}
\caption{The emissivity of the strongest individual emission lines for each ion as a function of temperature for $n_{\rm H} = 10^{-3}\,$cm\3, $z=0$, and solar abundances. Lines from the same ion are displayed with the same colour.
Lines are shown for hydrogen, helium, carbon, nitrogen, oxygen, neon and magnesium. }
\label{emissivity1}
\end{figure*}

\begin{figure*}
\centering
\includegraphics[width=0.9\textwidth]{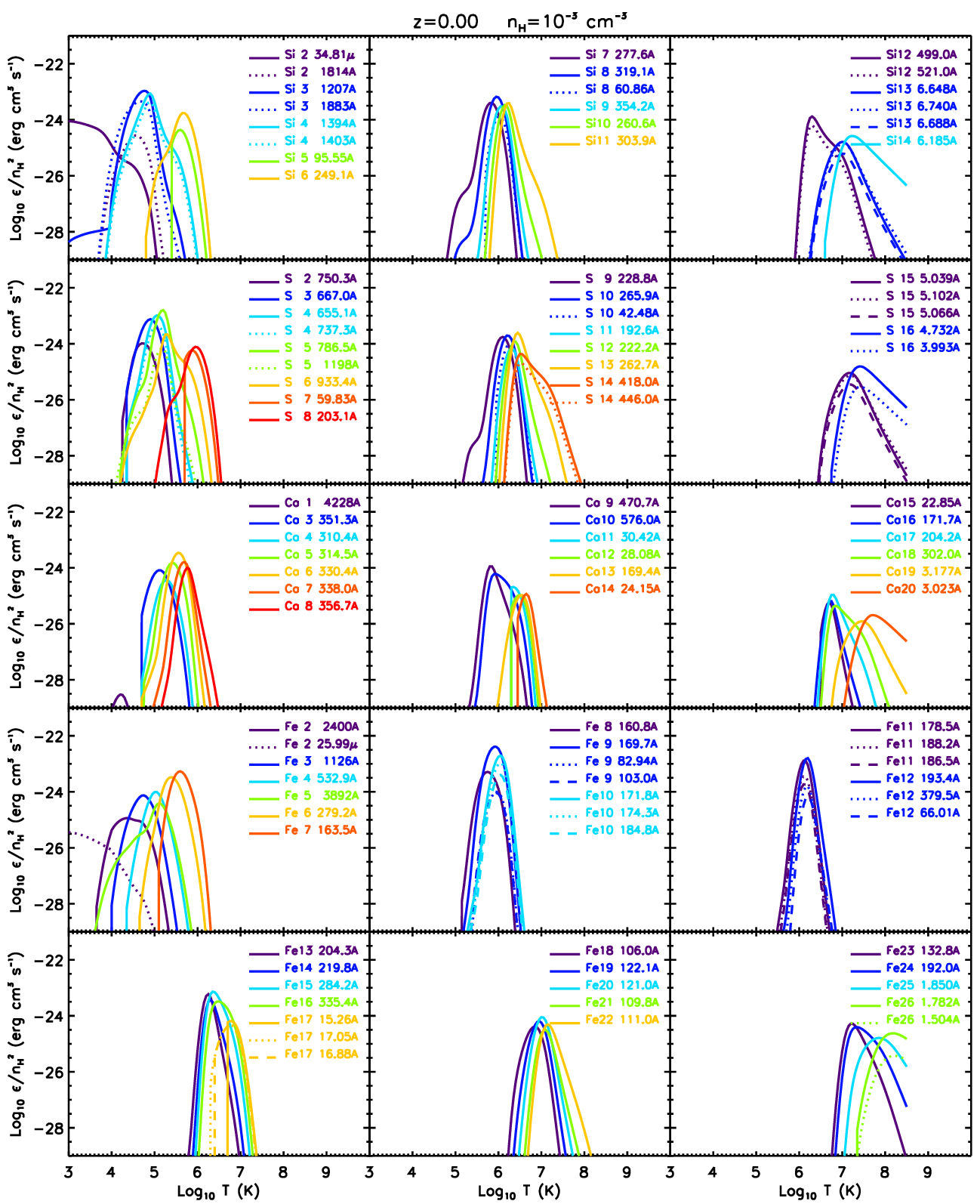}
\caption{As Fig.~\ref{emissivity1}, but for silicon, sulphur, calcium and iron.}
\label{emissivity2}
\end{figure*}

\begin{figure*}
\centering
\includegraphics[width=\textwidth]{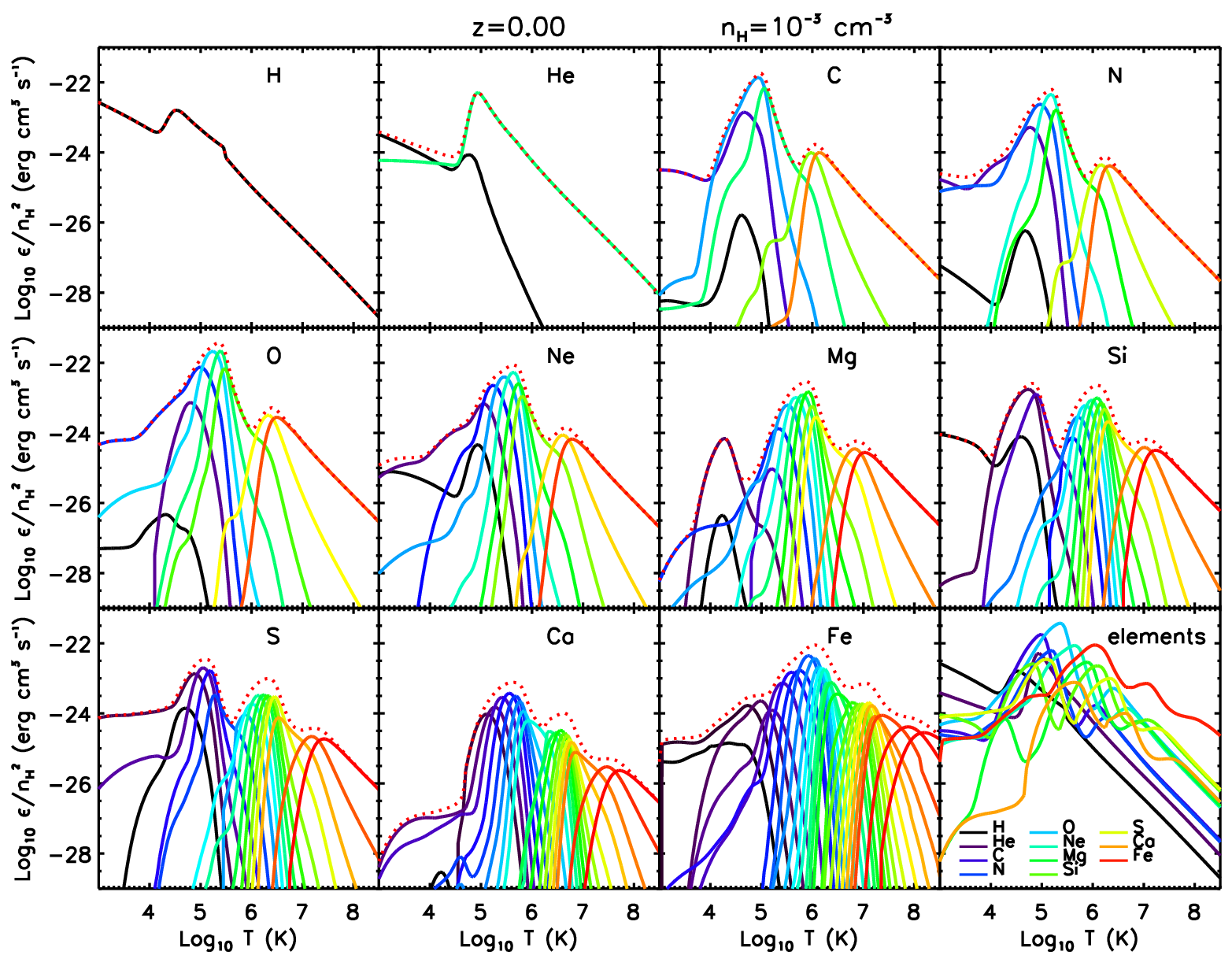}
\caption{The total line emissivity of ions as a function of temperature for $n_{\rm H} = 10^{-3}\,$cm\3, $z=0$, and solar metallicity. Each panel shows ions of a specific element. The line colours vary as violet-blue-green-yellow-orange-red with increasing ionic number. The dotted red line shows the total line emissivity of each element. The bottom-right panel shows the total line emissivities of each element.
The temperature for which the emissivity of each line is maximum increases with ionic number. With the exception of iron, the colour-coding of ions is the same as in Figs.~\ref{emissivity1} - \ref{emissivity2}.}
\label{ionz0}
\end{figure*}

We use the line emissivity tables created by \citet{bertone2010a}, which were also used in \citet{bertone2010b, bertone2011} and \citet{vandevoort2012}. They include a relatively small selection of the strongest lines for each ion -- slightly over 2000 emission lines, out of a total of several millions -- for the 11 elements explicitly tracked by \owls\  (H, He, C, N, O, Ne, Mg, Si, S, Ca and Fe). The contributions of all other elements are small, and should not significantly affect our estimate of the total emission. For comparison, calcium is the element in our sample that contributes the smallest fraction of the total energy, of the order of 0.1 per cent (see Section \ref{channel}). The contributions of the neglected elements are all smaller than this value \citep[see also][]{wiersma2009a}.

We restrict our attention to the lines with the highest emissivities for each ion, so that the total emissivity is accurate to within about a per cent or better. In many cases this requires just a few lines per ion (\civ\ 1548\AA, 1551\AA\ and \ovi\ 1032\AA, 1038\AA\ are good examples, carrying almost 100 per cent of the total energy emitted by the \civ\ and \ovi\ ions respectively), while for heavier atoms such as iron and calcium many more lines are required.

The emissivity of the strongest lines of each ion is shown as a function of temperature in Figs.~\ref{emissivity1} (H, He, C, N, O, Ne, and Mg) and \ref{emissivity2} (Si, S, Ca, and Fe). All emissivities are shown at $z=0$ for a constant hydrogen number density of $n_{\rm H}=10^{-3}\,$cm\3\ and solar element abundances. Note that emissivity tables are thus computed for a metallicity that substantially exceeds the average metallicity of diffuse gas, which is about $0.1\,Z_{\sun}$ at low redshift \citep{wiersma2009b}. 
For densities lower than the $n_{\rm H}=10^{-3}\,$cm\3\ assumed here, some of the emissivity curves show higher values in the low temperature regime, where photo-ionisation becomes dominant, as shown by \citet{bertone2010a}. At higher densities photo-ionisation becomes negligible, and the emissivity curves drop sharply for temperatures below the values corresponding to the peak emissivities, as is also the case for most metal lines for the density shown here. For clarity, for each ionic species we show at most the three lines with the highest peak emissivities, even if more lines are included in the tables. When less than three lines are shown, as for the UV doublets of lithium-like ions and for beryllium-like ions, the lines that are not shown have peak emissivities that are at least an order of magnitude smaller than the dominant lines. In some cases, as for most iron and calcium lines, we show only one line for each ion to avoid overcrowding the figures.

The total line emissivities of individual ions, calculated by summing up all the lines included in the tables for each ion, are shown in Fig.~\ref{ionz0} as a function of temperature, at redshift $z=0$, for $n_{\rm H} = 10^{-3}\,$cm\3\ and solar abundances.
The colour of the emissivity curves of each ion varies smoothly with increasing ionisation state; black-blue colours show the emissivities of lower ionisation states with low peak temperatures, while yellow-red colours represent highly ionised species, such as helium- and hydrogen-like ions. For these last two species, the emissivity peak roughly indicates the temperature threshold above which the atoms are nearly fully ionised. With the exception of iron lines, the colour-coding used for Figs.~\ref{emissivity1}-\ref{emissivity2} and Fig.~\ref{ionz0} is the same.

Fig.~\ref{ionz0} illustrates visually how the shape and the peak temperature of the line emissivity curves depend on the atomic number and on the ionisation state of each element. In particular, the peak temperature, i.e.\ the temperature for which the emissivity is maximum, increases steadily with increasing ionisation state and atomic number.
This can easily be understood, because more bound electrons require larger energies, and therefore higher temperatures, to be removed or excited.
Hydrogen and helium contain at most two electrons on the 1$s$ energy level that can be extracted at relatively low temperatures. Heavy elements with atomic numbers up to 11 (C, N, O, and Ne) have two electrons on the 2$s$ level and up to six electrons on the 2$p$ level; this translates into emissivity curve peaks that cluster in two separate temperature ranges: those for helium- and hydrogen-like ions at the highest temperatures, and all others for ions with remaining electrons on the 2$s$ or 2$p$ levels clustering at lower temperatures.
Elements with atomic numbers between 11 and 18 (which includes Mg, Si, and S) contain electrons on the 3$s$ and 3$p$ levels. These produce a further set of emissivity curves that cluster together and peak at the lowest temperatures. Finally, calcium and iron have electrons on the 4$s$ and 3$d$ levels respectively and for this reason a fourth and a fifth separate cluster of emissivity curves, corresponding to the upper energy levels, appear at the lowest temperatures.
Moving up the scale in atomic mass, each new set of emissivity curves, corresponding to a higher energy level, always appears at the lowest temperature, while the whole distribution moves to higher temperatures, with lines from hydrogen-like ions having the highest peak temperatures.

As we will show in the following Sections, and as already discussed by \citet{bertone2010a,bertone2010b}, the shapes and peak temperatures of the emissivity curves are fundamental in determining the level of the line emission. However, the relative intensity of the emissivity curves does not automatically translate into equal relative emission intensities, because the total emission also depends on the ionic abundance of different species and on the overall mass content in each temperature range. As an example, the emissivity of \lya\ lines is lower than those of many oxygen lines, but overall the total emission from hydrogen lines is much larger than that of oxygen lines, because hydrogen is much more abundant than oxygen. This will be shown in Section \ref{volume}.

\subsection{Continuum emissivity}
\label{cont_em}

In this Section we describe our technique to build the continuum emissivity tables, which is similar to that used for the line emissivity tables.

The total continuum emissivity can be separated into the contribution from gas purely composed of hydrogen and helium, and contributions from each of the metals tracked in \owls. The emissivity tables for heavy elements are computed using \cloudy, following the approach of \citet{wiersma2009a}. The emissivity due to metal $y$ is taken to be the difference between the emissivity computed assuming solar abundances and the emissivity computed after setting the abundance of $y$ to zero, while keeping the abundances of all other elements the same relative to hydrogen. As discussed by \citet{wiersma2009a}, this approach is accurate as long as the contribution of heavy elements to the free electron density is small, i.e.\ for metallicities $Z\la Z_\odot$. Since hydrogen and helium are both important contributors of free electrons, we tabulate the total emissivity for these elements combined and do so for a range of helium abundances (relative to hydrogen). 

Hence, while the line emissivity tables are 3-dimensional, the continuum tables are effectively 4-dimensional for metals and 5-dimensional for the combined hydrogen and helium continuum. As for the lines, the continuum emissivity is sampled as a function of density, temperature and redshift. One additional dimension is due to the wavelength dependence of the continuum, which is mapped over 12 orders of magnitude in wavelength in the range $\lambda = 0.01$ \AA\ to 1 m.
The tables for the contribution of hydrogen and helium include one more dimension, the 5th, that takes into account variations in the abundance of helium relative to that of hydrogen. Tables have been created for 8 different helium mass fractions, varying in the range $Y=0.238 - 0.308$ by bins of $\Delta Y=0.01$, with $Y=0.248$ being the primordial abundance.

\begin{figure}
\centering
\includegraphics[width=8.4cm]{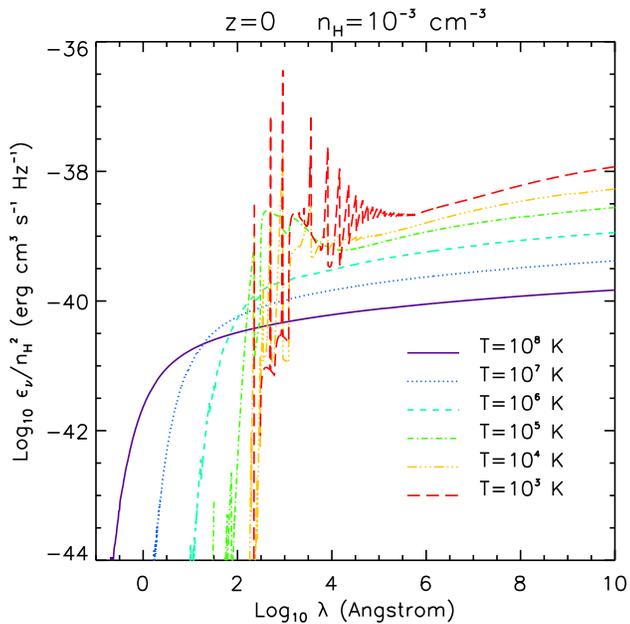}
\caption{The total continuum emissivity as a function of temperature for constant density, $n_{\rm H} = 10^{-3}\,$cm\3, $z=0$, and solar abundances. The overall shapes of the curves reflect the wavelength and temperature dependence of thermal Bremsstrahlung. The peaks at intermediate wavelengths that become prominent for low temperatures correspond to radiative recombinations and 2-photon emission from hydrogen and helium.}
\label{contemp}
\end{figure}

The key factor determining the level and the shape of the continuum emissivity is the gas temperature. In Fig.~\ref{contemp} we show how the total continuum emissivity for a gas with solar composition varies as a function of temperature between $T=10^3\,$K and $T=10^8\,$K. The overall shape of the curves reflects the wavelength and temperature dependence of the emissivity for thermal Bremssstrahlung (i.e.\ free-free emission), which scales as $T^{-1/2}\exp(-hc/\lambda kT)$, where $h$ is Planck's constant, $c$ is the speed of light and $k$ is Boltzmann's constant. For $T < 10^6\,$K the emissivity curve becomes less smooth at intermediate wavelengths as radiative recombinations and two-photon decays of metastable levels become increasingly important for lower temperatures (e.g.\ \citealt{osterbrock1974}; \citealt{sutherland1993}; \citealt{ferland1998}). From left-to-right (i.e.\ from short to long wavelengths), the sharp peaks in the $10^3\,$K curve correspond to the \heii\ Lyman limit (228\AA, $n=1$), the \hei\ Lyman limit (504\AA, $n=1$), and the \hi\ Lyman (912\AA, $n=1$), Balmer (3646\AA, $n=2$), and Paschen (8204\AA, $n=3$) limits. The sharp peaks at longer wavelengths correspond to recombinations to higher energy levels ($n > 3$). The broader peak at $\approx 2431$\AA\ corresponds to 2-photon emission from \hi. 

We do not show the contributions of the individual elements to the continuum emissivity as we find that it is nearly always dominated by hydrogen and helium. The exception is the regime $hc/\lambda \gg kT$; however in this regime the continuum emissivity is exponentially suppressed. Finally, we note that dust emission, which we have ignored here as it is not clear how much dust resides in the IGM, is expected to become increasingly important at longer wavelengths.

\section{Simulations}
\label{owlproj}

\begin{figure*}
\centering
\includegraphics[width=0.8\textwidth]{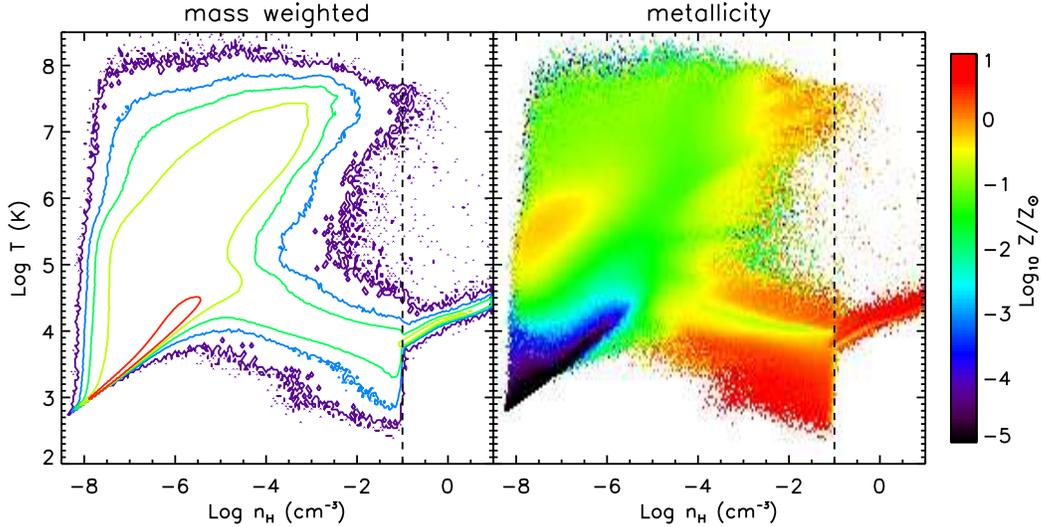}
\caption{The mass-weighted distribution of gas in the temperature-density plane (left panel, contours are spaced by 1 dex) and the mass-weighted average gas metallicity (right panel) for the $REF\_L100N512$ run at $z=0$. The vertical, dashed lines indicate the star formation threshold, above which we impose an effective equation of state onto the gas. The gas metallicity is not a good tracer of the gas mass, but is highly biased towards high densities.}
\label{massmet}
\end{figure*}

In this Section we briefly describe the suite of cosmological, hydrodynamical simulations that constitute the \owls\ project \citep{schaye2010}, with particular attention to the model referred to as the ``reference'', or \default, model.

The \owls\ project aims to investigate the effects of varying a large number of physical prescriptions on the predicted evolution of the baryons in the Universe. At present, more than fifty simulations have been run, with varying mass resolutions, box sizes, and physical prescriptions for gas cooling, star formation, chemical evolution, supernova and AGN feedback. The simulations have been run with a custom version of {\sc gadget III}, a significantly extended version of the parallel PMTree Smoothed Particle Hydrodynamics (SPH) code {\sc gadget II} \citep{springel2005}.

Our basic results have been computed using two simulations of the reference model, which contain $2\times 512^3$ particles each. The first, labelled $REF\_L025N512$, has a box size of 25 \hm\ comoving Mpc and follows the evolution of structures down to $z=2$. The gas and dark matter particle masses are\footnote{These are the initial values. The masses of baryonic particles vary slightly during the course of the simulation as a result of stellar mass loss.} $1.4\times 10^6$ \hm\ \msun\ and $6.3\times 10^6$ \hm\ \msun, respectively, and the gravitational force resolution (i.e.\ the softening) is 0.5 \hm\ comoving kpc, but is frozen at 0.5 \hm\ proper kpc below $z=2.91$. The second simulation, denoted by $REF\_L100N512$, has a box size of 100 \hm\ comoving Mpc and follows the evolution of structures down to $z=0$. Compared with $REF\_L025N512$, this simulation uses $4^3=64$ times larger particle masses and a 4 times larger softening.
We have tested additional runs with variations in the physical prescriptions, box sizes and resolution, which we describe in more detail in Section \ref{sims} and in the Appendix.

Besides gravitational and hydrodynamical forces, the simulations employ new physical prescriptions for radiative cooling, star formation, chemodynamics, supernova and AGN feedback.
Radiative cooling is implemented following the prescription of \citet{wiersma2009a}. The cooling rates are calculated for each element under the assumption of an optically thin gas in ionisation equilibrium, in the presence of the cosmic microwave background and an ionising background radiation from quasars and galaxies \citep{haardt2001}. Pre-computed tables for each element have been created with \cloudy\footnote{version 05.07 rather than 07.02 as in \citet{wiersma2009a}.}, as described for the continuum emissivity in Section~\ref{cont_em}. The total gas cooling rate is calculated by interpolating the tables\footnote{Using equation (3) rather than (4) of \citet{wiersma2009a}.} and by summing up the contributions of each element, rescaled to the particle element abundance.

Star formation is implemented as described in \citet{schaye2008}. Gas with densities $n_{\rm H} > 10^{-1}\,$cm\3\ is expected to be multi-phase and unstable to star formation. As we lack both the physics and the resolution to model the multiphase interstellar medium, we impose a polytropic equation of state onto these particles of the form $P\propto \rho^{4/3}$. This choice of power-law index results in a constant Jeans mass, which prevents numerical fragmentation. The observed Kennicutt-Schmidt star formation law is analytically converted into a pressure law and gas particles with $n_{\rm H} > 10^{-1}\,$cm\3\ are stochastically converted into star particles at the pressure-dependent rate implied by this law. As shown by \citet{schaye2010} and \citet{haas2012}, the star formation rates are insensitive to the assumed star formation law as a result of self-regulation. We will not consider any simulations with variations in the star formation prescriptions.
The prescription for chemical evolution is described in \citet{wiersma2009b}. The scheme tracks the evolution of the abundance of the 11 elements that contribute significantly to the radiative cooling rate (H, He, C, N, O, Ne, Mg, Si, S, Ca, and Fe) produced by massive stars (core collapse supernovae and stellar winds) and intermediate mass stars (Type Ia supernovae and asymptotic giant branch [AGB] stars), assuming a \citet{chabrier2003} stellar initial mass function.

Feedback from core collapse supernovae and stellar winds from massive stars are included as kinetic feedback following the prescription of \citet{vecchia2008}. The model describes winds using two free parameters, the mass-loading factor $\eta=\dot{m}_{\rm w}/\dot{m}_*$, which parametrizes the rate at which mass is injected into the wind, $\dot{m}_{\rm w}$, as a function of the star formation rate $\dot{m}_*$, and the wind initial velocity, $v_{\rm w}$. The \default\ model assumes $\eta=2$ and $v_{\rm w}=600$ \kms, which correspond to about 40 per cent of the energy available from core collapse supernovae. Model \default\ does not include AGN feedback.

The mass-weighted distribution of gas in the temperature-density plane is shown in the left panel of Fig.~\ref{massmet} for the $REF\_L100N512$ run at $z=0$, while the right panel shows the mass-weighted average gas metallicity. The distributions for the $REF\_L025N512$ run (not shown) are qualitatively similar, but the gas overall fills a more compact region of the plane. In particular, the maximum gas temperatures at $z=2$ are lower than at $z=0$ by about half a dex, and the minimum densities are higher by a similar factor.

At both redshifts, the bulk of the metals do not trace the bulk of the IGM mass.
Instead, most of the gas mass is concentrated along the temperature-density relation corresponding to the low-density, photo-ionised IGM (red contour in the left panel of Fig.~\ref{massmet}), while metal-rich gas concentrates in the circum-galactic medium, with $n_{\rm H}>10^{-5}\,$cm\3\ and $T<10^5\,$K, and in the intra-group and intra-cluster media, with $n_{\rm H}>10^{-5}\,$cm\3\ and $T>10^6\,$K. A significant amount of metals can finally be found in the low-density, warm-hot IGM with $10^5\,$K $<T<10^6\,$K. A more in depth study of the metal distribution in these runs can be found in \citet{wiersma2009b}.

\subsection{The gas emission}
\label{gasem}

The gas line emission is estimated following the methodology of \citet{bertone2010a}. The luminosity of each particle is calculated by interpolating the emissivity tables as a function of density and temperature at the desired redshift and is then rescaled to the corresponding element abundance.

The luminosity $L_{l,i}\left(z\right)$ of particle $i$ corresponding to line $l$ of element $y$ at redshift $z$, in units of erg s\1, is calculated as
\begin{equation}\label{lumin}
L_{l,i}\left(z\right) = \varepsilon_{l, \odot} \left( z, T_i,n_{{\rm H},i} \right) \frac{m_{{\rm gas},i}}{\rho_i} \frac{X_{{\rm y},i}}{X_{\rm y \sun}},
\end{equation}
where $m_{{\rm gas},i}$ is the gas mass and $\rho_i$ the density of particle $i$. $\varepsilon_{l,\odot} \left( z, T,n_{\rm H} \right)$ is the line emissivity in units of erg cm\3\ s\1\ and is bi-linearly interpolated (in logarithmic space) from the \cloudy\ tables as a function of the particle temperature $T_i$ and hydrogen number density $n_{{\rm H},i}$.
$X_{{\rm y},i}$ is the mass fraction of element $y$ and $X_{\rm y \sun}$ its solar value. For our calculations we use the ``smoothed'' element abundances described in \citet{wiersma2009b}, which are defined as the ratio of the SPH smoothed metal mass density and the SPH smoothed mass density, $Z_{{\rm y},i} = \rho_{{\rm y},i} / \rho_{i}$. This definition of the gas metallicity is more consistent with the SPH formalism than the commonly used particle metallicity, i.e.\ the ratio of the metal and total mass of the particle. As discussed in \citet{wiersma2009b}, the use of smoothed abundances has the additional advantage that it reduces the effects of the lack of metal mixing that is inherent to the SPH technique.

Similarly to the line emission, the continuum emission is calculated by interpolating the tables to the desired density, temperature, and redshift. The tables for the contributions of the metals are interpolated individually and rescaled to the corresponding elemental abundances. The hydrogen$+$helium tables are in addition interpolated to the helium abundance of the particles.
As a result, the continuum luminosity as a function of wavelength and redshift
$L_{c,y,i}\left(z,\lambda \right)$ for metal $y$ and particle $i$, in units of erg s\1\ Hz\1, is given by:
\begin{equation}\label{luminc}
L_{c,y,i}\left(z,\lambda \right) = \epsilon_{y,\lambda,\odot} \left( z, T_i, n_{{\rm H},i} \right) \frac{m_{{\rm gas},i}}{\rho_i} \frac{X_{{\rm y},i}}{X_{\rm y \sun}},
\end{equation}
with $\epsilon_{y,\lambda,\odot} \left( z, T,n_{\rm H} \right)$ the continuum emissivity of metal $y$ in units of erg cm\3\ s\1\ Hz\1. The continuum emission for hydrogen$+$helium is
\begin{equation}\label{luminh}
L_{c,{\rm H+He},i}\left(z,\lambda \right) = \epsilon_{{\rm H+He},\lambda} \left( z, T_i, n_{{\rm H},i}, X_{{\rm He},i} \right) \frac{m_{{\rm gas},i}}{\rho_i},
\end{equation}
with $\epsilon_{{\rm H+He},\lambda} \left( z, T,n_{\rm H}, X_{\rm He}  \right)$ the combined continuum emissivity for hydrogen and helium, also in units of erg cm\3\ s\1\ Hz\1. The total continuum emissivity of particle $i$ is then
\begin{equation}\label{totc}
L_{c,i}\left(z,\lambda \right) = L_{c,{\rm H+He},i}\left(z,\lambda \right) + \sum_{y} L_{c,y,i}\left(z,\lambda \right).
\end{equation}

We will often plot the volume-weighted, specific emission $E_{\rm v}$, which is also an average emissivity, computed by summing the luminosities of all gas particles in the simulation and dividing by the simulation volume. We will compute this specific emission for selected lines and continuum components, for the sum of all lines of particular ions, and for the sum of all emission processes from all ions. Finally, if we only include gas particles that fall in a certain temperature range, we can compute these quantities as a function of the gas temperature.

\section{The cooling channels of diffuse gas}
\label{volume}

In this Section we present results for the two reference simulations: \default\_L025N512 at $z=2$ and \default\_L100N512 at $z=0$. In Section~\ref{channel} we discuss the relative contributions of ions and the continuum to the global emission budget, and in Section~\ref{continuum} we investigate the main gas cooling channels as a function of temperature. Finally, in Section \ref{global} we present results for the global energy density of the diffuse IGM.

\subsection{Cooling channels integrated over all diffuse gas}
\label{channel}

\begin{figure*}
\centering
\includegraphics[width=0.8\textwidth]{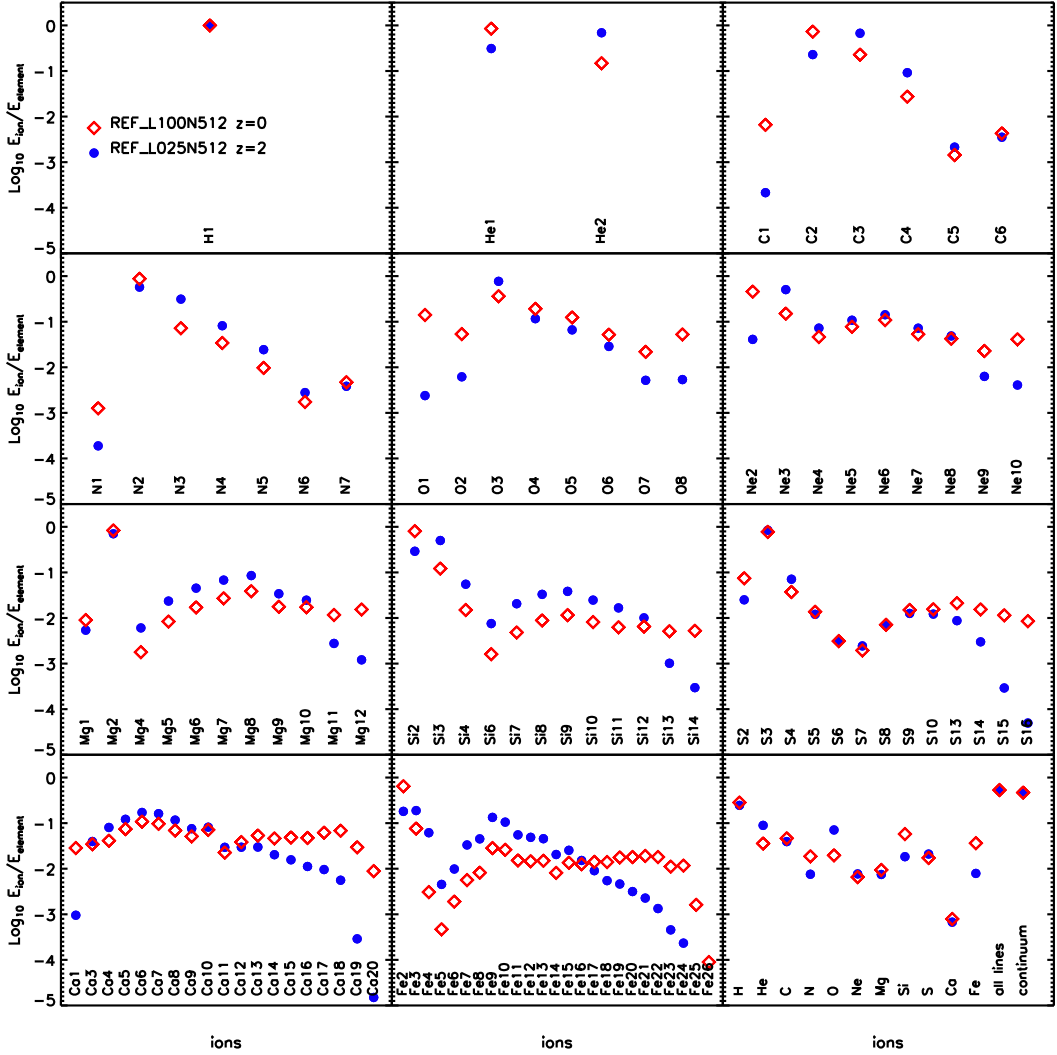}
\caption{Fraction of the energy emitted per unit volume by (almost) all ions for each element, averaged over all the diffuse gas in the simulation. In each panel a different element is shown and the sum of the contributions of all ions corresponding to that element equals 1. The bottom right panel shows the fraction of energy per unit volume emitted by each element, by all lines and by continuum, with respect to the total specific emission. Results are for the \default\_L100N512 simulation at $z=0$ and for the \default\_L025N512 simulation at $z=2$.}
\label{fracions_100}
\end{figure*}

\begin{figure*}
\centering
\includegraphics[width=0.49\textwidth]{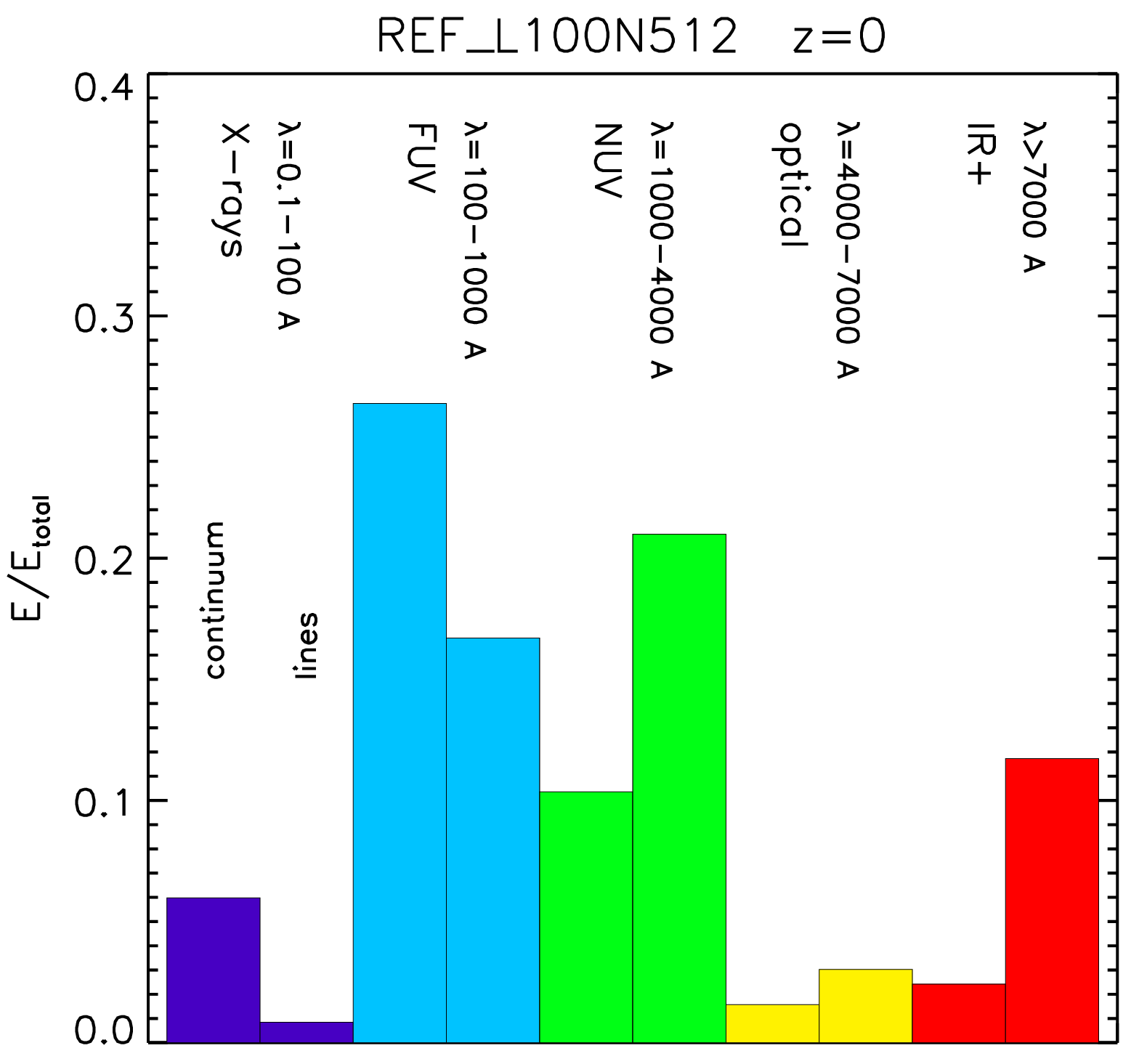}
\includegraphics[width=0.49\textwidth]{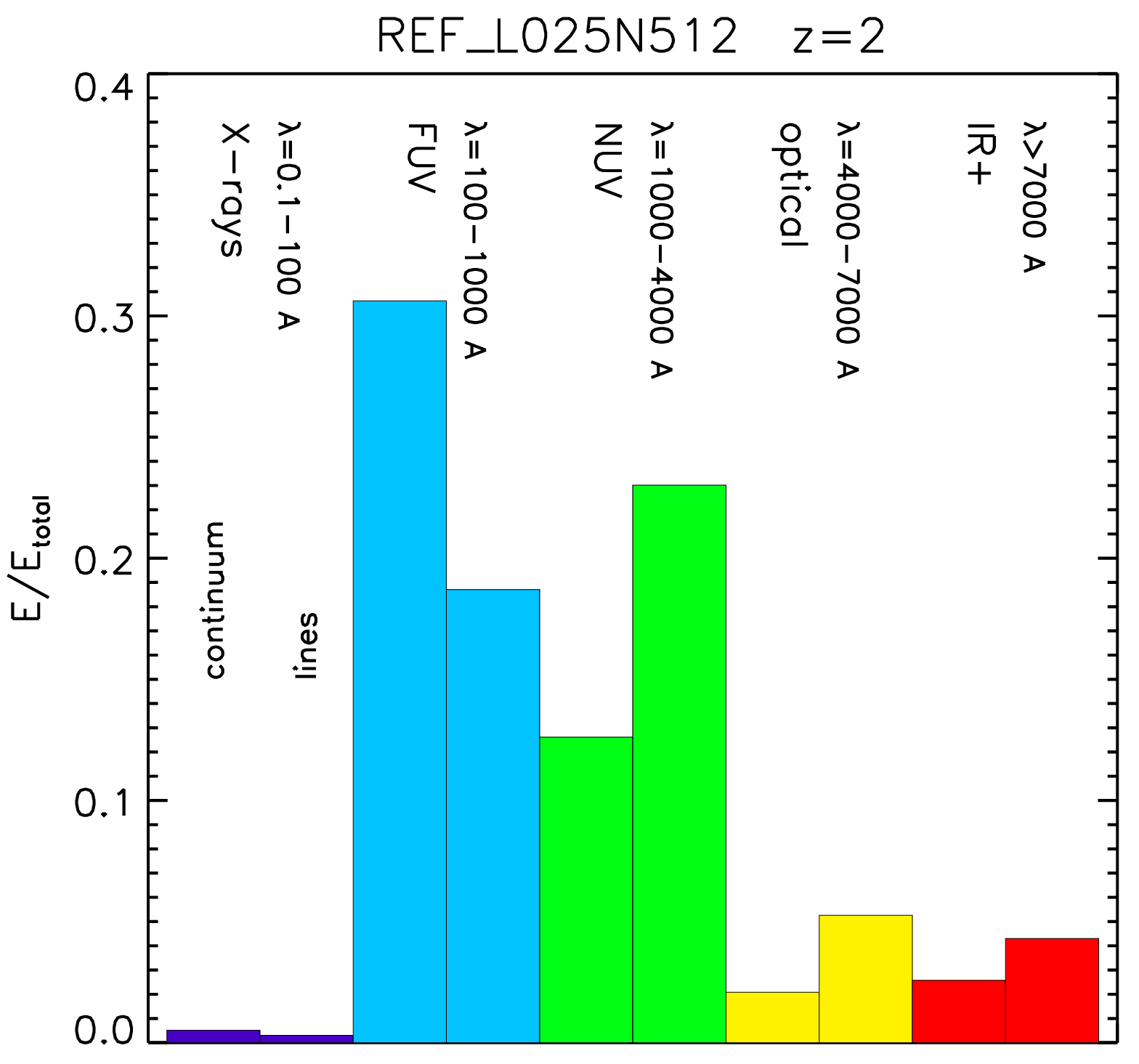}
\caption{Fraction of the energy emitted in the continuum emission in different wavebands, averaged over all the diffuse gas in the simulation. Results are for the \default\_L025N512 simulation at $z=2$ (left panel) and for the \default\_L100N512 simulation at $z=0$ (right panel).
In both cases, more than half the energy is emitted in the FUV band, and a large fraction of the remaining energy is emitted in the NUV band. The \xray\ band becomes significant only at low redshift, where it emits more energy than the optical and IR bands combined. Here, the IR band includes all wavelengths larger than 7000 \AA\ (i.e., including the proper IR and all other bands up to radio frequencies).}
\label{cont_frac}
\end{figure*}

In this Section we discuss the relative contributions of different ions and the continuum to the global emission budget of diffuse gas, integrated over the entire simulation volume.

Fig.~\ref{fracions_100} shows results for the fraction of energy emitted in the form of lines by each ion\footnote{The emissivity tables do not include any lines for a few ionic species, because their emissivities are very small. The neglected ions are \nei, \mgiii, \sili, \silv, \si, \sxi, \sxii, \caii\ and \fei. The corresponding curves are thus missing from Figs.~\ref{ions_temp100} and \ref{ions_temp25}. \label{missing_ions}}, relative to the total energy in lines carried by the corresponding element, for all elements explicitly followed in the \owls\ runs. The total line energy emitted by all ions in each panel equals one.
Results are shown at $z=0$ for \default\_L100N512 (red diamonds) and at $z=2$ for \default\_L025N512 (blue, filled circles). The bottom right panel of Fig.~\ref{fracions_100} shows the fraction of energy emitted in the form of lines by each element, with the last two columns showing the total fraction in lines and continuum. 
This panel demonstrates that lines and continuum carry similar amounts of energy, with lines being slightly more important. Of the emision lines, those from hydrogen carry most of the energy, their contribution to the total emission being about thirty per cent. At $z=0$ the next most important elements are silicon, carbon, helium, and iron, whose lines contribute $\approx 3-6$ per cent each. Next come oxygen, nitrogen, sulphur, magnesium, and neon which contribute $\approx 0.6-2$ per cent each. Finally, calcium lines only account for about 0.08 per cent of the energy. 

Looking now at the other panels, we see that there are a few common features to all elements and ions. Most of the energy in lines is emitted by atoms in relatively low ionisation states, which indicates that dense, warm gas is responsible for much of the energy dissipated in the form of line radiation. Lines from neutral gas in
general contribute a minor fraction of the emission, with the notable
exception of \oi, which contributes about 15 per cent of the energy
emitted by oxygen at $z=0$ (but is unimportant at $z=2$).  

At $z=2$ the relative importance of lines and continuum is unchanged, but there are significant changes in the contributions from individual elements. Helium and oxygen are much more important and each account for nearly ten per cent of the energy. This comes at the expense of silicon, nitrogen, and iron which account for about $3-7$ times less energy than at $z=0$. The fact that helium lines contribute less at lower redshift can be understood by noting that the metallicity of the diffuse gas increases with time, providing a larger range of cooling channels in the relevant temperature range. The change in relative contributions of the heavy elements is also partly a consequence of the chemical evolution of the diffuse gas. While core collapse supernovae dominate the metal production rate early on, intermediate mass stars release large amounts of nitrogen and iron on longer time scales through AGB winds and type Ia supernovae \citep[e.g.][]{wiersma2009b}.

The evolution in the relative contributions of individual ions is larger than that for elements. The lowest ionisation states are much less important at higher redshifts, probably because there is less dense gas and because the UV background is more intense. The highest ionisation stages of oxygen and heavier elements carry much smaller fractions of the energy at higher redshifts. This is
because at high redshift the fraction of hot, metal enriched gas is
very small compared to the mass of warm gas (see
e.g.\ \citealt{cen1999}; \citealt{wiersma2009b}).

\begin{table} 
\begin{center} 
\caption{Wavelength bands.}
\label{table_bands}
\begin{tabular}{lrr}
\hline
Band & $\lambda_{\rm min}$ (\AA) & $\lambda_{\rm max}$ (\AA)\\
\hline
\xray & 0.1 & 100\\
Far ultraviolet (FUV) & 100.0 & 1000\\
Near ultraviolet (NUV) & 1000.0 & 4000\\
Optical & 4000.0 & 7000\\
Infrared and longer (IR+) & 7000.0 & 1 m\\
\hline
\end{tabular}
\end{center}
\end{table}

Fig.~\ref{cont_frac} shows how the fraction of energy emitted through continuum radiation and lines is distributed in the different wavelength bands defined in Table~\ref{table_bands}. The left panel shows results for \default\_L100N512 at $z=0$ and the right panel for \default\_L025N512 at $z=2$.

Independent of redshift, almost half the energy is emitted in the FUV band and another third in the NUV band. As such, about 80 per cent of the total energy is emitted in the UV wavelength range $\lambda = 100 - 4000$ \AA. The optical emission contributes 5-8 per cent of the energy, and the IR+ band about 7-15 per cent. The band with the largest variation with redshift is the \xray\ band: its contribution increases from less than 1 per cent at $z=2$ to about 7 per cent at $z=0$.

The \default\_L025N512 simulation has a box size of only 25 \hm\ Mpc and it is possible that the fraction of energy emitted in the \xray\ band is underestimated because the most massive haloes are not sampled. However, since the fraction of hot gas with $T>10^8\,$K does not exceed a few per cent in mass at $z=0$, we do not expect to have underestimated the contribution of the \xray\ band by more than a few per cent. 

Overall, the continuum emission dominates the energy budget at the highest energies (\xray\ and FUV bands), while emission lines are dominant at longer wavelengths ($\lambda > 1000$ \AA).
The relative contributions of lines and continuum vary little with redshift. The exceptions are lines with $\lambda > 7000$ \AA, whose contribution increases from 4 to 12 per cent of the total over time, tripling their share in the IR+ band.

\begin{figure}
\centering
\includegraphics[width=8.4cm]{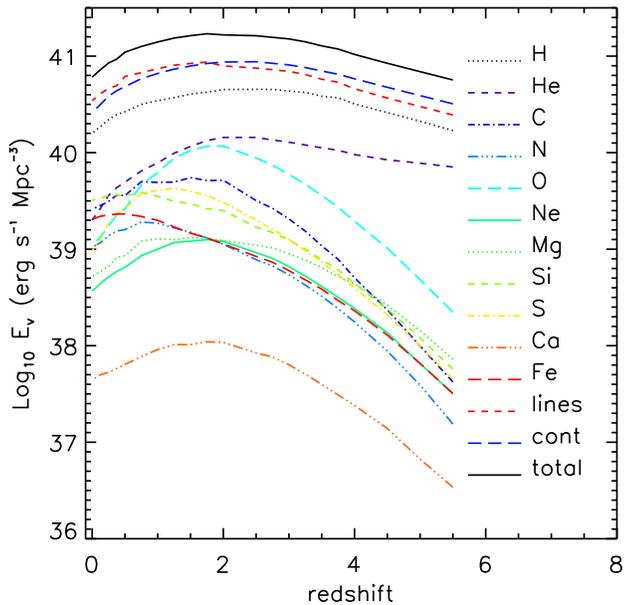}
\caption{Evolution of the specific emission (emission per unit volume) of lines from individual elements and from the continuum, averaged over all the diffuse gas. Results are for the \default\_L100N512 simulation.}
\label{redshift}
\end{figure}

The evolution of the sums of the volume-averaged comoving emissivities of all emission lines from individual elements, along with the total radiative cooling rates in the forms of lines and continuum, are shown for $z=0-5.5$ in Fig.~\ref{redshift} for simulation \default\_L100N512. The global emissivity increases by about a factor of 3 from $z=5.5$ to 2, after which it drops increasingly rapidly to a value that is similar to the $z=5.5$ value. The shape of the curve is very similar to that of the cosmic star formation history in this model \citep[see][]{schaye2010}, consistent with the idea that the rate at which diffuse gas cools ultimately driving the rate at which galaxies form stars. As for the star formation rate, the evolution is driven by structure formation, which increases the density contrast and thus the cooling rates (most of which scale as the density squared), and the general expansion of the Universe, which decreases the gas densities and slows down the formation of structure, particularly when the energy density of the Universe ceases to be dominated by matter.

Results for the \default\_L025N512 (not shown) and \default\_L100N512 simulations converge well below $z\approx 4$. At higher redshifts, the \default\_L025N512 simulation predicts significantly higher global emissivities. The high-redshift divergence of these two simulations, which again parallels that of the cosmic star formation rates \citep{schaye2010}, occurs because the higher-resolution run can resolve lower mass galaxies that form earlier than galaxies in the \default\_L100N512 run and that can enrich the IGM with
metals at higher redshifts. The greater fraction of high-density gas and the higher metallicity both increase the global, high-redshift cooling rate.

Continuum and line emission each account for about half the emission over the whole redshift range, with lines becoming relatively more important with time. Hydrogen accounts for about half of the line emission at all redshifts. While helium is by far the second most dominant element at high redshift, for $z<2$ oxygen is responsible for  a similar amount of line emission. At $z<1$ the contributions of all elements heavier than hydrogen are within a factor of three of each other, except for neon, magnesium, and particularly calcium, which are less important in terms of the global emission budget. 

The relative strength of the emission from different elements varies with time as a consequence of chemical evolution, evolution in the distribution of mass in the temperature-density plane, and evolution in the UV background radiation. The elements that differ the most from the general trend are iron and silicon, whose global emissivity continues to increase down to $z\sim 0.5$. For iron this is probably due to a combination of two facts: the increase of the fraction of hot gas, where iron emission is most efficient, and the increase in iron produced by type Ia supernovae, which release their nucleo-synthetic products on much longer time scales than elements such as oxygen and neon that are predominantly released by massive stars. The production of silicon is, however, also dominated by massive stars. To understand its evolution, we note that at low redshift its emission is totally dominated by the \silii\ 34.8$\mu$ line, which is produced by cold $T\sim 10^3\,$K gas (see Figs.~\ref{fracions_100} and \ref{lines_vs_temp_z=0}). Hence, it is the red cloud of high-metallicity, high-density, low-temperature gas in the right panel of Fig.~\ref{massmet} that dominates the emission. At high redshift the gas has a lower metallicity and less of it is able to cool down to the temperatures where \silii\ becomes important.

\subsection{Cooling channels as a function of temperature}
\label{continuum}

\begin{figure*}
\centering
\includegraphics[width=\textwidth]{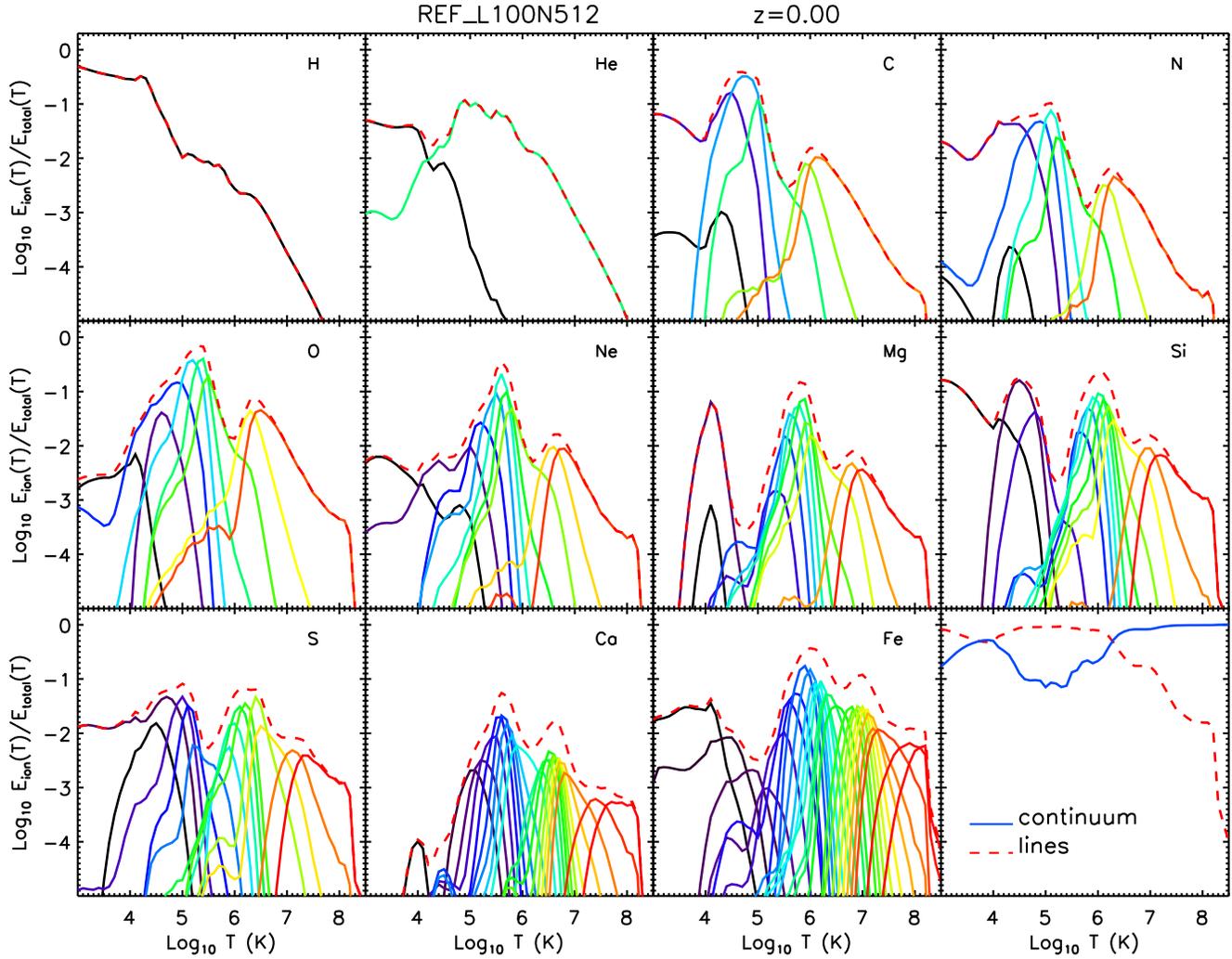}
\caption{Fraction of the energy emitted per unit volume in the form of lines as a function of temperature, averaged over all the diffuse gas in the \default\_L100N512 simulation at $z=0$. The different curves show different ions and the different panels show different elements, except for the bottom-right panel, which compares the fractions of the total mean emission per unit volme due to lines (red, dashed) and continuum (solid, blue). The fractions
are relative to the total emission in each temperature bin. The results were obtained by averaging the emission over the entire volume of the \default\_L100N512 simulation at $z=0$. The curves end at the maximum temperature sampled by the particles in the simulation volume.}
\label{ions_temp100}
\end{figure*}

\begin{figure*}
\centering
\includegraphics[width=\textwidth]{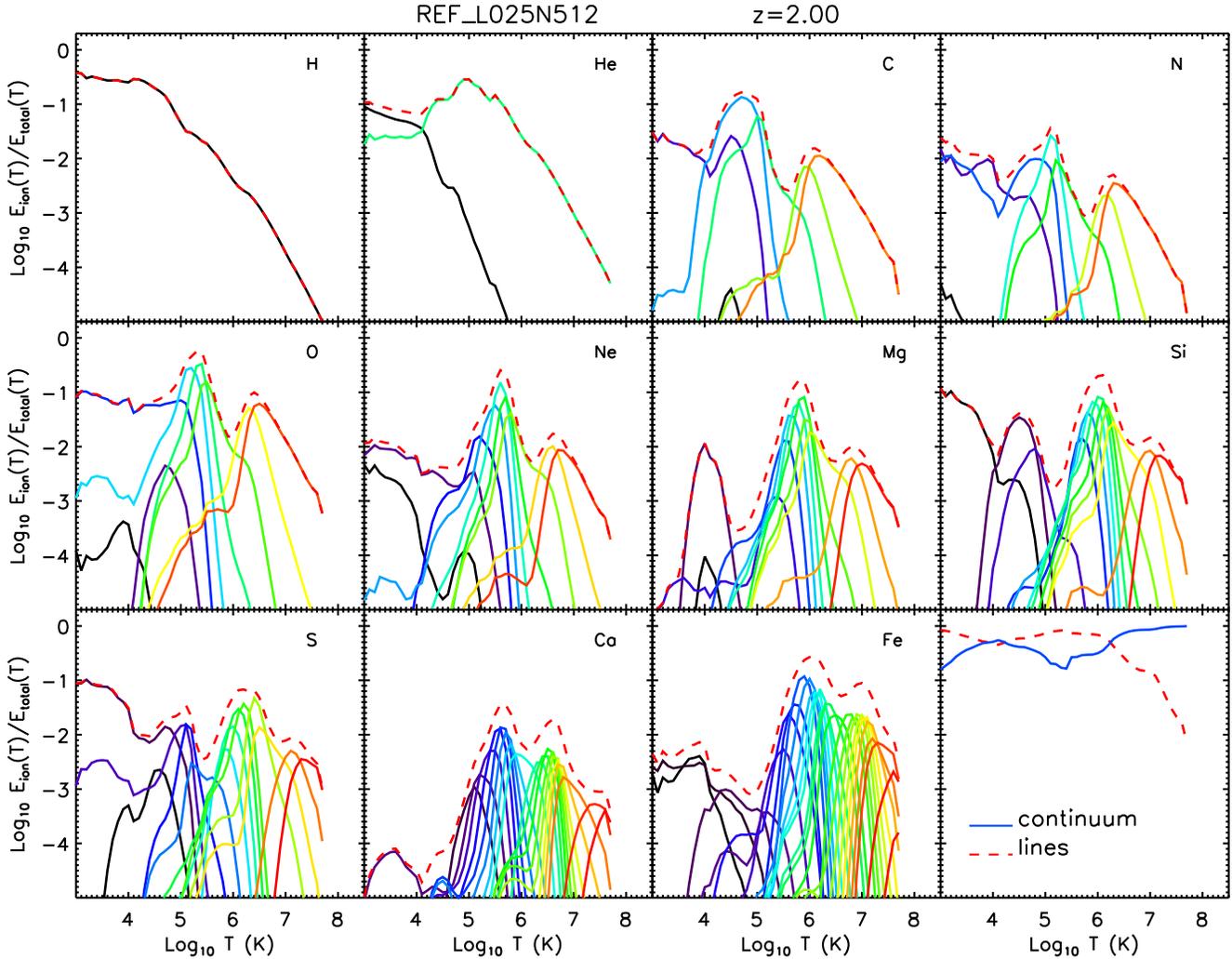}
\caption{As Fig.~\ref{ions_temp100}, but for the \default\_L025N512
simulation at $z=2$.}
\label{ions_temp25}
\end{figure*}

In this Section we discuss the main cooling channels of diffuse gas as a function of temperature.

Fig.~\ref{ions_temp100} shows the relative contributions of the total line emission from individual ions to the average energy that is emitted per unit time and volume as a function of the gas temperature. The results are for \default\_L100N512 at $z=0$ and have been averaged over all the diffuse gas in the entire simulation volume. Different panels correspond to different elements, as indicated, and different coloured curves correspond to different ions. For each ion, all lines have been summed. The curves are normalised separately for each temperature, i.e.\ the solid curves show the fraction of the total emission from gas at that particular temperature emitted by a single ion. The dashed red curve in each panel shows the sum of all the ions of the corresponding element, i.e., it is the sum of all the solid curves in the same panel. Finally, the dashed red and the solid blue curves in the bottom-right panel show the fractions of the total energy that is emitted in the form of lines and continuum, respectively, summed over all elements. The two curves therefore sum to unity. 

Fig.~\ref{ions_temp25} is similar to Fig.~\ref{ions_temp100} but shows the results for \default\_L025N512 at $z=2$. The two simulations differ in terms of their resolution and box size. However, as the results are nearly converged with respect to both resolution and box size (see the Appendix), they can be directly compared in order to check for evolution. 

It is important to note that Figs.~\ref{ions_temp100} and \ref{ions_temp25} do not simply reflect the emissivity curves. As the emission is averaged over the entire simulation volume, the results depend on the distribution of mass and the chemical composition of the gas as a function of temperature and density. The figures thus illustrate the relative contributions of line and continuum radiation (bottom-right panel) and the relative contributions of different elements and ions to the cooling rates of diffuse gas as a function of temperature after averaging over the entire (simulated) universe.

Focusing first on the bottom-right panel of Fig.~\ref{ions_temp100}, we see that emission lines carry most of the energy at temperatures lower than a few times $10^6\,$K, while continuum emission is dominant for higher temperatures, where it can exceed the contribution of lines by more than two orders of magnitude. This is because most elements are very highly ionised at such high temperatures, thereby severely reducing the number of bound electrons that can be collisionally excited. Bremsstrahlung therefore dominates the cooling rates of such high-temperature plasmas. Line and continuum radiation contribute approximately equally to the cooling rates for $T\sim 10^4\,$K, but lines dominate for both lower and higher temperatures. At $T\sim 10^5\,$K the continuum only accounts for about ten per cent of the radiative losses.

Fig.~\ref{ions_temp100} shows that among emission lines, \hi\ lines carry most of the energy for $T\la 10^4\,$K, while metal lines are dominant for $T > 10^5\,$K. All elements that we consider contribute fractions $\gg 1$ per cent of the total energy in some temperature range. 
The elements (ions) whose lines contribute substantially more than ten per cent of the total energy are hydrogen for $T\la 10^4\,$K (\hi), carbon at $\sim 10^{4.7}\,$K (\cii\ and \ciii), oxygen for $\sim 10^5\,$K (\oiii, \oiv, \ov, and \ovi), neon at $10^{5.6}\,$K (\nev), magnesium for $10^{5.9}\,$K, silicon at $\sim 10^3\,$K (\sili), $\approx 10^{4.5}\,$K (\silii), and $\sim 10^6\,$K, and iron for $\sim 10^6\,$K (\feviii\ and \feix) and at $T \approx 10^7\,$K. 

Comparing Figs.~\ref{ions_temp100} and \ref{ions_temp25}, we find little evolution in the distributions, with the main difference being the relative contributions of lines and the continuum in the range $10^{4.5}<T<10^6$ K. In this temperature interval, the contribution of lines is about twice as large at $z=0$ than at $z=2$. The increase is due to the higher average metallicity of the IGM at low redshift.

\begin{figure*}
\centering
\includegraphics[width=\textwidth]{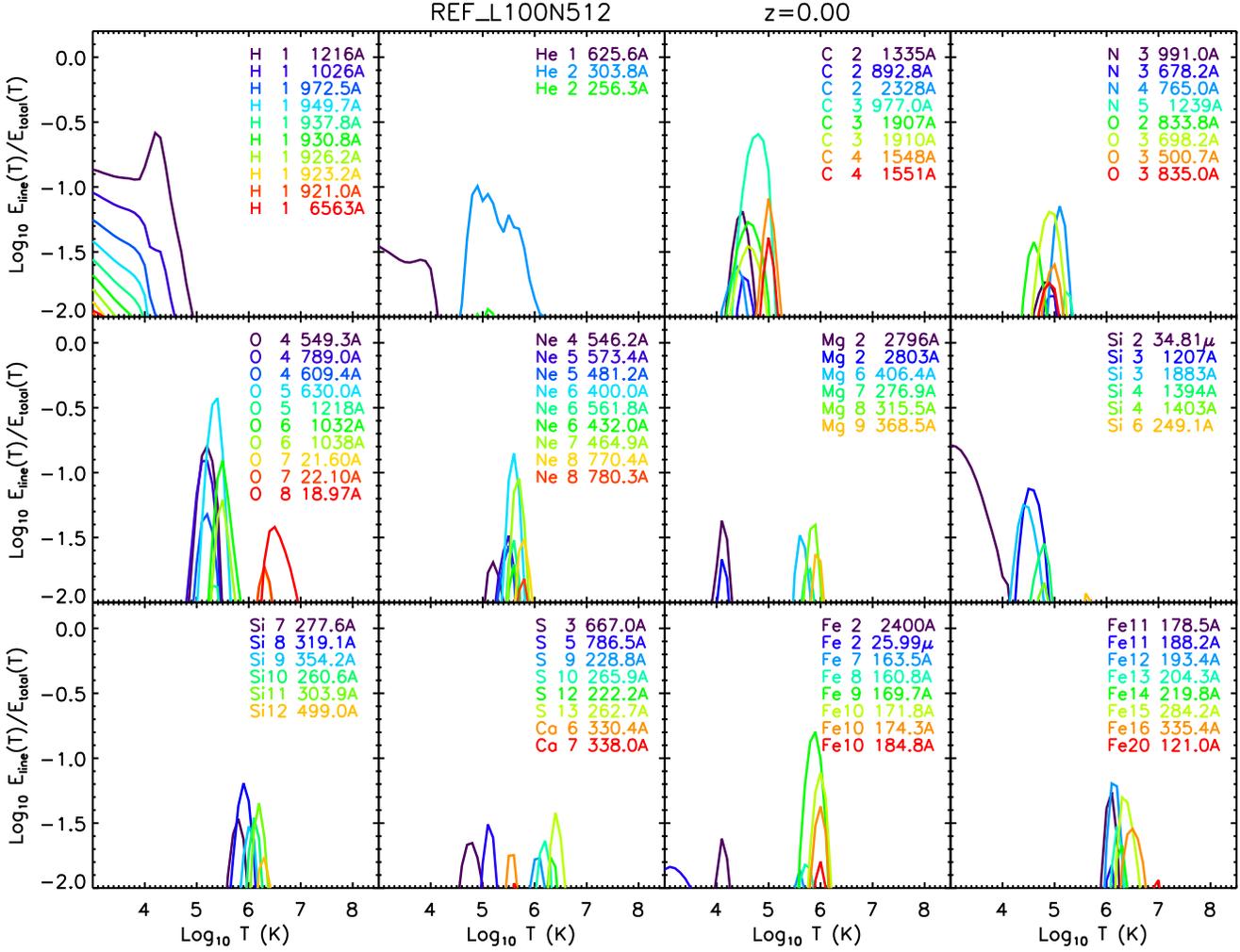}
\caption{Fraction of the energy emitted per unit volume in the form of individual lines as a function of temperature. The fractions are relative to the total emission in each temperature bin. Only lines that contribute at least 1 per cent of the total emitted energy in at least one temperature bin are shown. The results were obtained by averaging the emission of diffuse gas over the entire volume of the \default\_L100N512 simulation at $z=0$.}
\label{lines_vs_temp_z=0}
\end{figure*}

\begin{figure*}
\centering
\includegraphics[width=\textwidth]{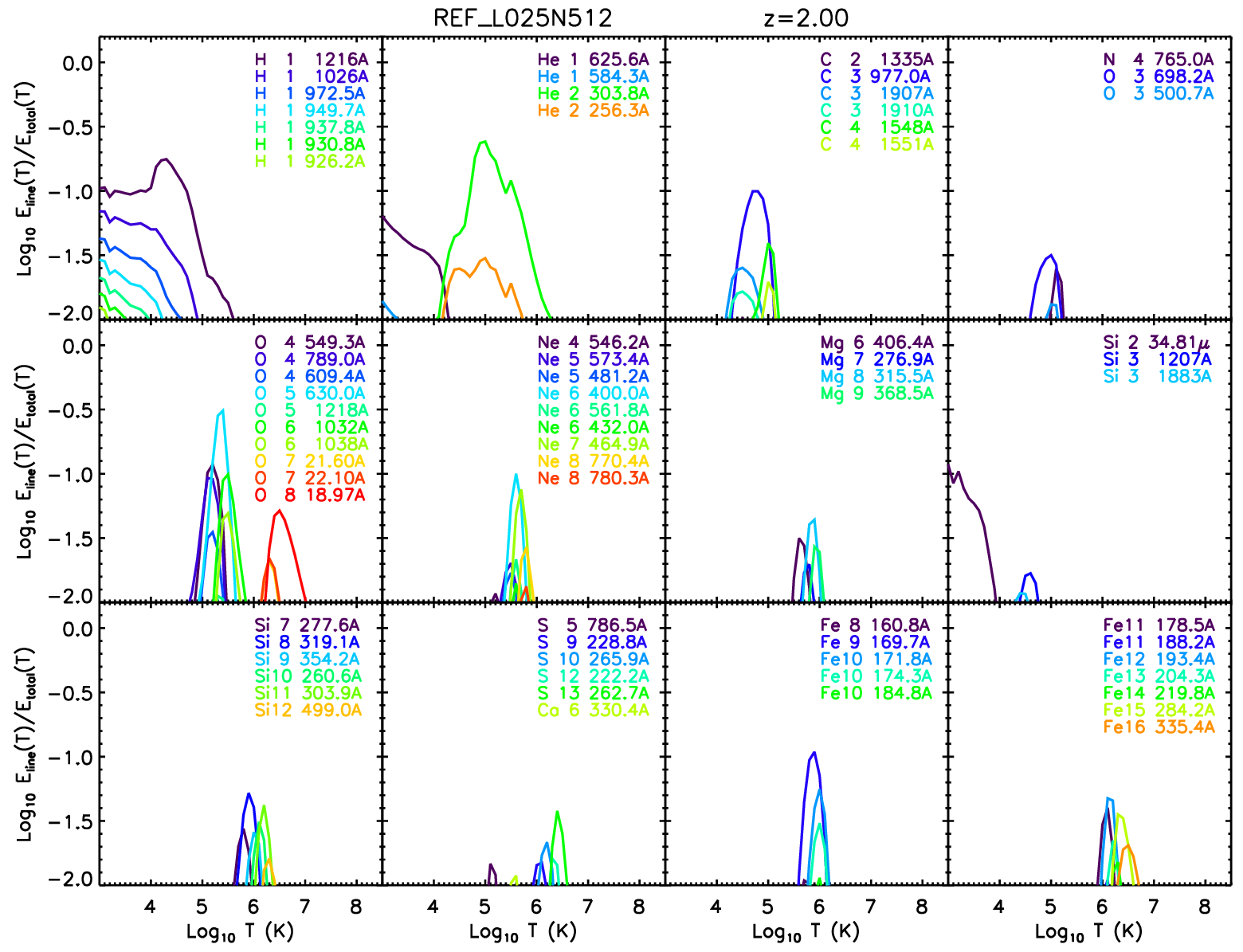}
\caption{As Fig.~\ref{lines_vs_temp_z=0}, but for the \default\_L025N512
simulation at $z=2$.}
\label{lines_vs_temp_z=2}
\end{figure*}

Figs.~\ref{lines_vs_temp_z=0} and \ref{lines_vs_temp_z=2} show the contributions of individual lines, rather than ions, at $z=0$ and $z=2$ respectively. All lines that contribute more than 1 per cent of the total emitted energy in at least one temperature bin are shown and they are listed in Tables \ref{linelist1} and \ref{linelist2} in Appendix \ref{linelist}. The Tables show the total fraction of energy radiated through each line along with the fraction of energy carried by the line at the temperature for which the line is most important. Both quantities have been sorted by the fraction of the energy that they carry.

The lines that account for more than ten per cent of the total energy are \hi\ 1216\AA\ for $T\la 10^4\,$K, \heii\ 304\AA\ at $\approx 10^{4.9}\,$K, \ciii\ 977\AA\ at $\sim 10^{4.8}\,$K, \oiv\ 549\AA\ at $\approx 10^{5.2}\,$K, \ov\ 630\AA\ at $\approx 10^{5.4}\,$K, \ovi\ 1032\AA\ at $10^{5.5}\,$K, \nevi\ 400\AA\ at $10^{5.5}\,$K, \silii\ 34.8$\mu$ at $\sim 10^3\,$K, and \feix\ 170\AA\ at $\approx 10^{5.9}\,$K. 
Comparing Figs.~\ref{lines_vs_temp_z=0} and \ref{lines_vs_temp_z=2}, which show $z=0$ and 2, respectively, we see that the same lines dominate, but that most of the metal lines are less important at higher redshift and that the reverse is true for the \heii\ lines.

\begin{figure*}
\centering
\includegraphics[width=0.49\textwidth]{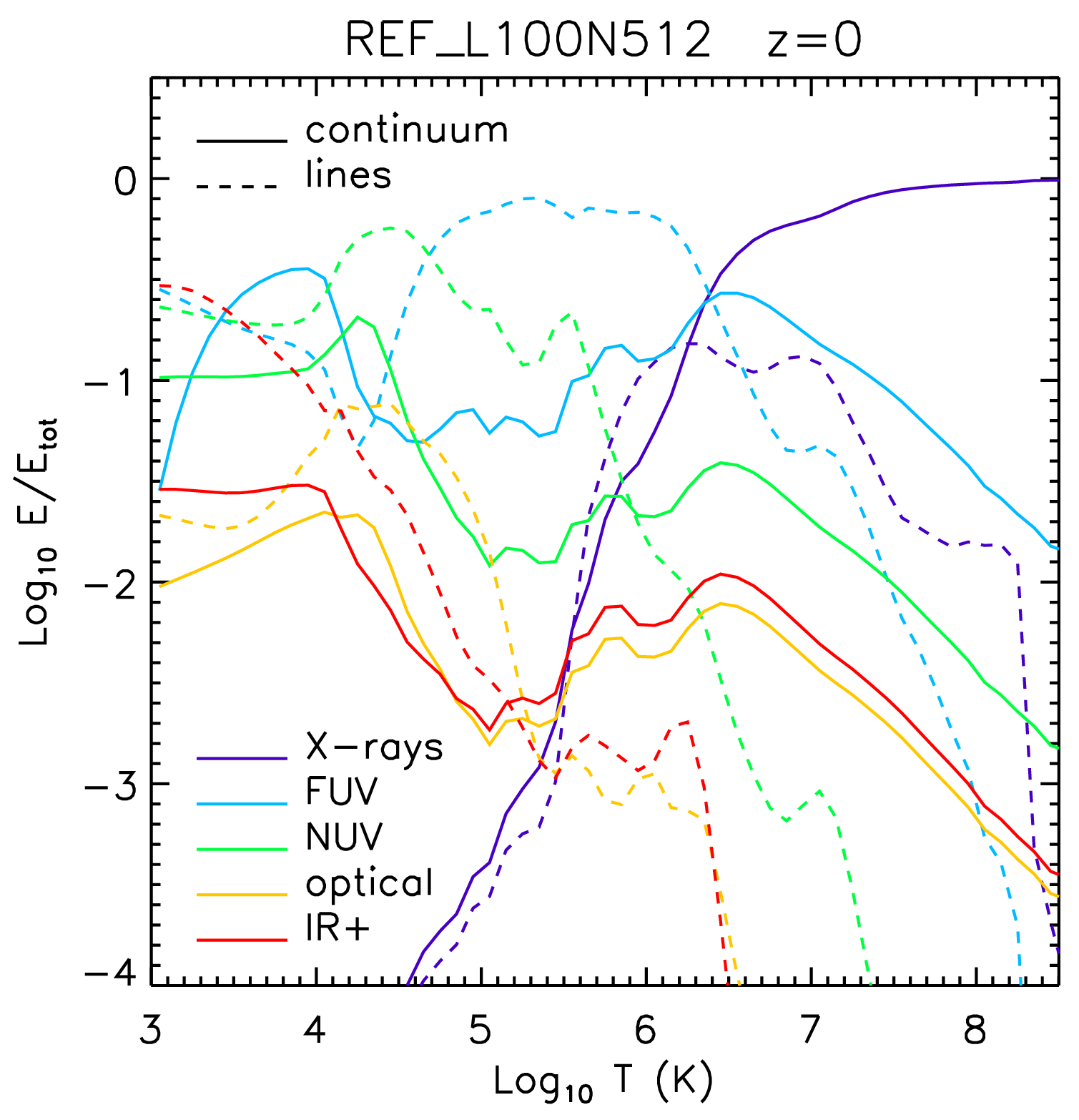}
\includegraphics[width=0.49\textwidth]{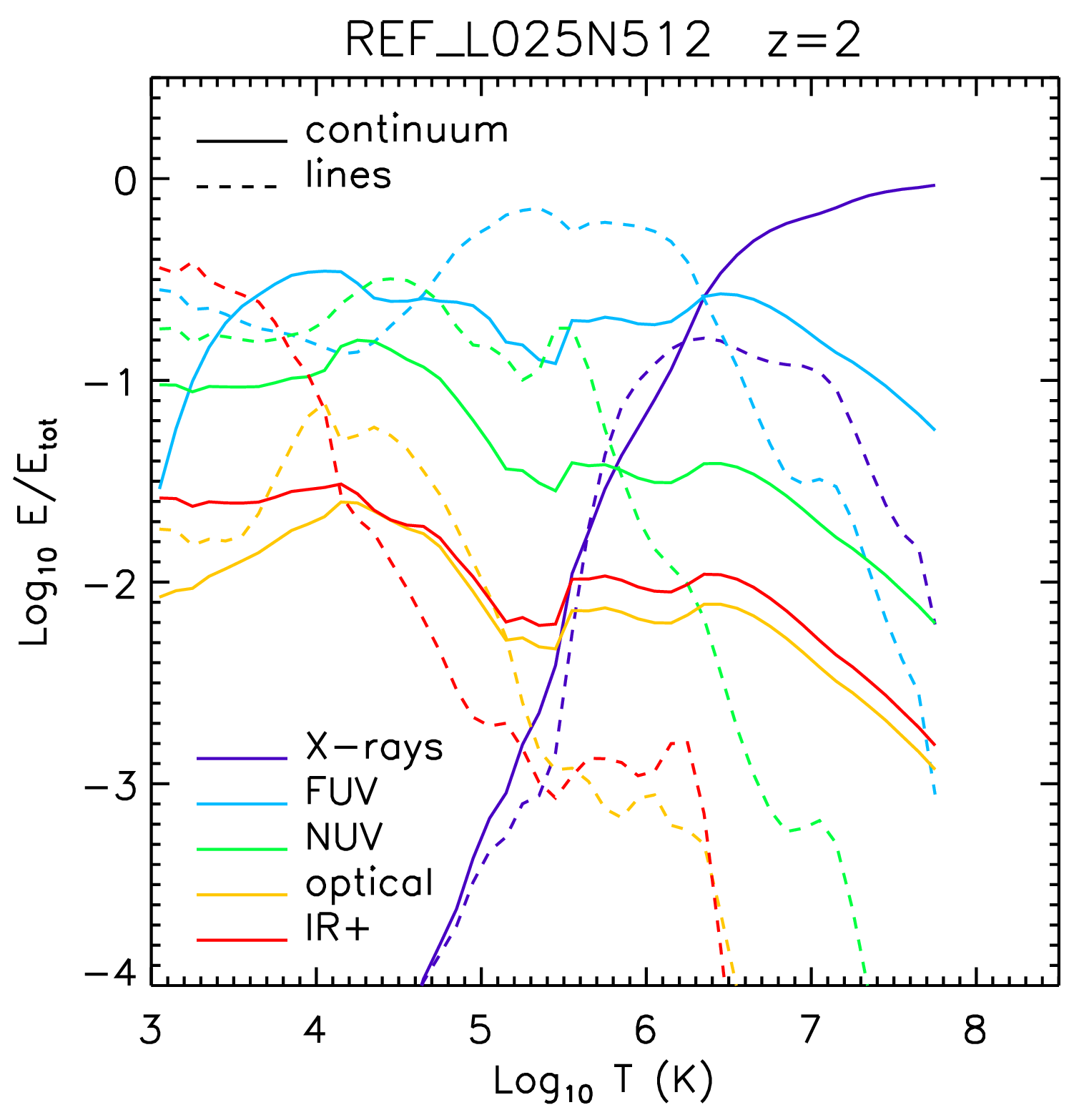}
\caption{The mean fraction of the total energy emitted by diffuse gas per unit volume accounted for by emission lines (dashed) and continuum radiation (solid) and by different wavelength bands (colours; see Table \ref{table_bands} for definitions). The results are averaged over the entire simulation volume. The left and right panels are for \default\_L100N512 at $z=0$ and \default\_L025N512 at $z=2$, respectively. For $T>10^{6.5}\,$K most of the energy is emitted as continuum radiation in the \xray\ band, while UV emission lines dominate the radiative cooling rate at $10^4<T<10^{6.5}$. For $T\sim 10^3\,$K the sums of all emission lines falling in the IR+, FUV, and NUV bands each account for about 30 per cent of the radiative energy losses. There is little evolution.}
\label{onlycont}
\end{figure*}

Fig.~\ref{onlycont} shows how, for each temperature, the fraction of energy emitted through lines and continuum radiation is distributed over the wavelength bands defined in Table~\ref{table_bands}. The left and right panels show results for \default\_L100N512 at $z=0$ and \default\_L025N512 at $z=2$, respectively. Note that the sum of all wavelength bins yields the
curves shown in the bottom-right panels of Figs.~\ref{ions_temp100}
and \ref{ions_temp25}. 

As the temperature increases, a greater fraction of the energy is emitted at shorter wavelengths. For $T \gg 10^6\,$K the \xray\ continuum is by far the most important cooling channel, while UV lines account for the vast majority of radiative losses for $10^4\,$K $ \ll T \la 10^6\,$K. If we sum line and continuum radiation, then the UV dominates down to the lowest temperature plotted, $T=10^3\,$K. IR+ lines only become important at $T\sim 10^3\,$K, where they account for about 30 per cent of the emitted energy. The maximum contribution of the optical band is about 10 per cent and is reached near $10^4\,$K, with most of the energy coming out in the form of lines. Comparing the two panels, we see little evolution, although the FUV continuum is more important for gas with $T\sim 10^5\,$K at high redshift.

\subsection{The global energy density of the IGM}
\label{global}

\begin{figure*}
\centering
\includegraphics[width=0.49\textwidth]{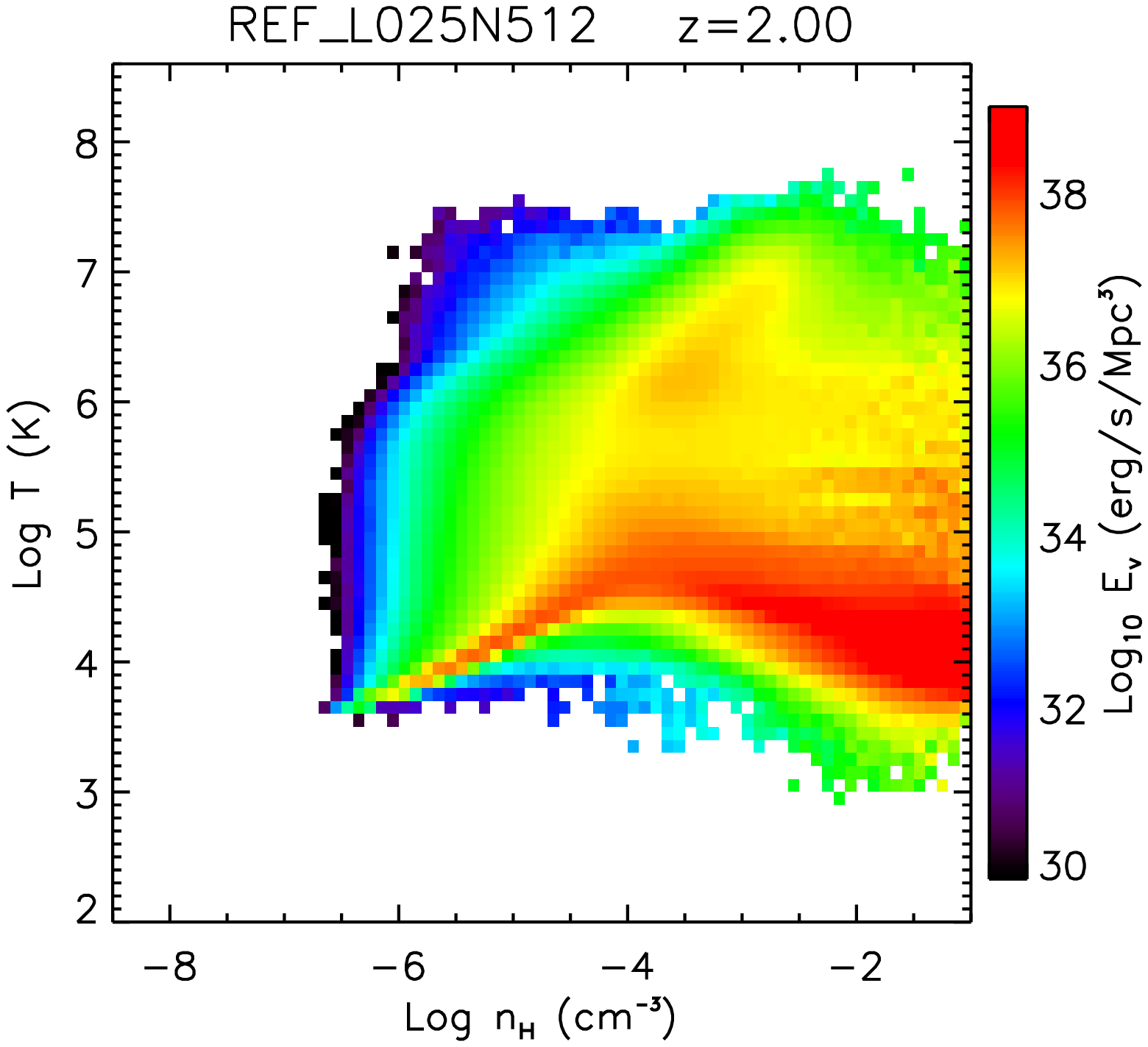}
\includegraphics[width=0.49\textwidth]{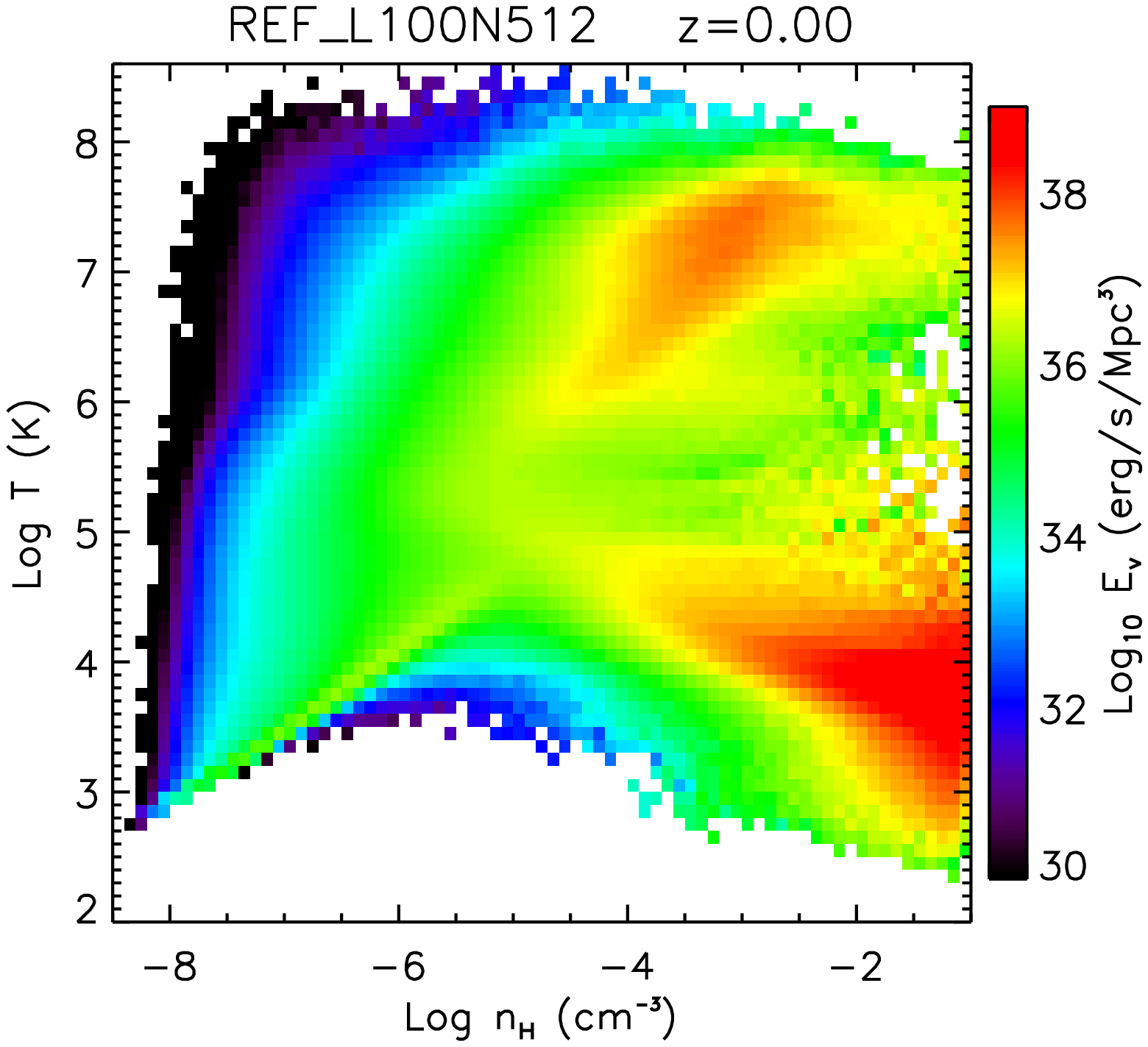}
\caption{The total emitted energy per unit volume as a function of the gas density and temperature. The left panel shows results for the \default\_L100N512 simulation at $z=0$, the right panel for the \default\_L025N512 at $z=2$. Most of the energy is emitted by relatively dense and metal rich gas.}
\label{totenergy}
\end{figure*}

Finally, we estimate the importance of the cooling radiation emitted by diffuse gas to the cosmic energy budget of the Universe.
According to \citet{fukugita2004}, the contribution of the IGM to the cosmic energy budget due to kinetic energy associated to the gas peculiar velocities is of order $\Omega_{\rm IGM,kinetic}\sim 10^{-8}$. Indeed, we find $\Omega_{\rm IGM,kinetic} \approx 5\times 10^{-8}$ for \default\_L100N512 at $z=0$ and $\approx 0.9\times 10^{-8}$ for \default\_L025N512 at $z=2$. These values are comparable to the total internal energy, for which we measure $\Omega_{\rm IGM,thermal} \approx 3\times 10^{-8}$ at both redshifts. Low density gas with $\rho/\rho_{\textrm{mean}} < 10$ carries about half the internal energy at $z=2$ and about 10 per cent at $z=0$.

The fact that the total kinetic and internal energy densities are comparable is not surprising. In regions where the kinetic energy density exceeds the thermal one, peculiar velocities will typically be supersonic, resulting in the conversion of kinetic to thermal energy through shocks. If, on the other hand, the thermal energy density is substantially larger than the kinetic energy, e.g.\ as a resuls of explosive feedback, then the gas will tend to expand at the velocity of sound until this velocity becomes comparable to the local escape velocity, which will in turn be similar to the typical peculiar velocities in the region.

By dividing the cosmic thermal energy density by the global cooling rate, we infer a cosmic cooling time of about 4.1~Gyr at $z=0$ and 1.4~Gyr at $z=2$.
For both redshifts the radiative cooling times are about a factor of three smaller than the age of the Universe. Thus, the diffuse IGM as a whole would tend to radiate a large fraction of its internal energy over a Hubble time, if all heating sources were removed. The fact that the radiative cooling time is comparable to the expansion time scale probably reflects the fact that structure formation, and therefore heating through accretion shocks and feedback from star formation, proceeds on this time scale and that radiative cooling and shock-heating approximately match each other.

Figure \ref{totenergy} shows the total emitted energy per unit volume as a function of the gas density and temperature. The left panel shows results for the \default\_L100N512 simulation at $z=0$, the right panel for the \default\_L025N512 at $z=2$. The total emitted energy in each density and temperature bin has been calculated as the sum of the continuum emission, integrated over all wavelengths, plus the energy emitted by all lines. When compared to Fig. \ref{massmet}, Fig. \ref{totenergy} demonstrates that the bulk of the emitted energy traces neither the bulk of the mass nor the bulk of the metals in the simulations. Instead, Fig. \ref{totenergy} can be better understood by comparing it to Fig. 13 of \citet{bertone2010a}, Fig. 10 of \citet{bertone2010b} and Fig. 9 of \citet{bertone2011}. These figures show the temperature-density distributions of gas, weighted by the emission in different UV and \xray\ lines, and indicate how the bulk of the emission is produced by the denser gas, independently of the gas mass distribution in the case of the metal lime emission. Similarly, Fig. \ref{massmet} shows that most of the total emission at both redshifts is produced by relatively dense gas, in accordance with the $\propto \rho^2$ dependence in the cooling law. The main difference between the results at $z=0$ and at $z=2$ is the increasing importance of emission from warm-hot and hot gas ($T>10^6$ K) with decreasing redshift, as seen also in Figs. \ref{fracions_100} and \ref{cont_frac}.

\section{Simulations with varying physics}
\label{sims}

\begin{figure*}
\centering
\includegraphics[width=\textwidth]{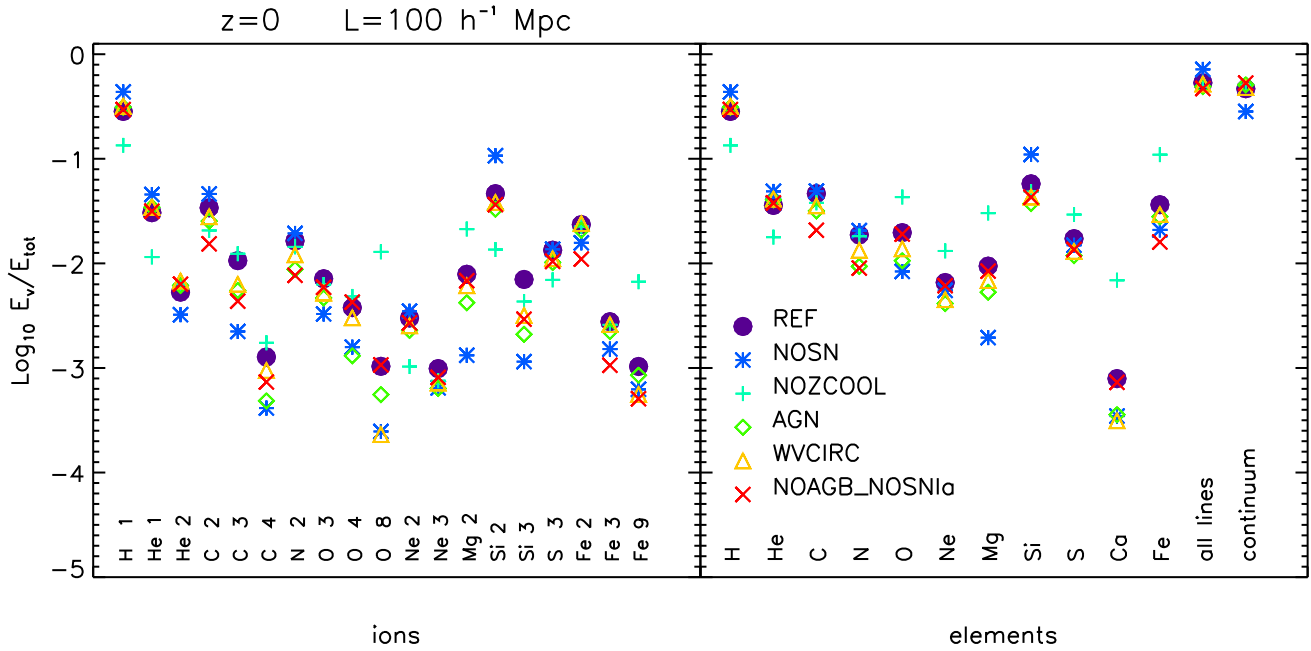}
\caption{The specific emission (emission per unit volume) at $z=0$ for a subset of \owls\ runs with boxes of 100 \hm\ Mpc on a side. The left panel shows the ions with the largest contributions to the total emission, while the right panel shows the contributions of individual elements, all lines combined and the continuum.}
\label{sims_100}
\end{figure*}

\begin{figure*}
\centering
\includegraphics[width=\textwidth]{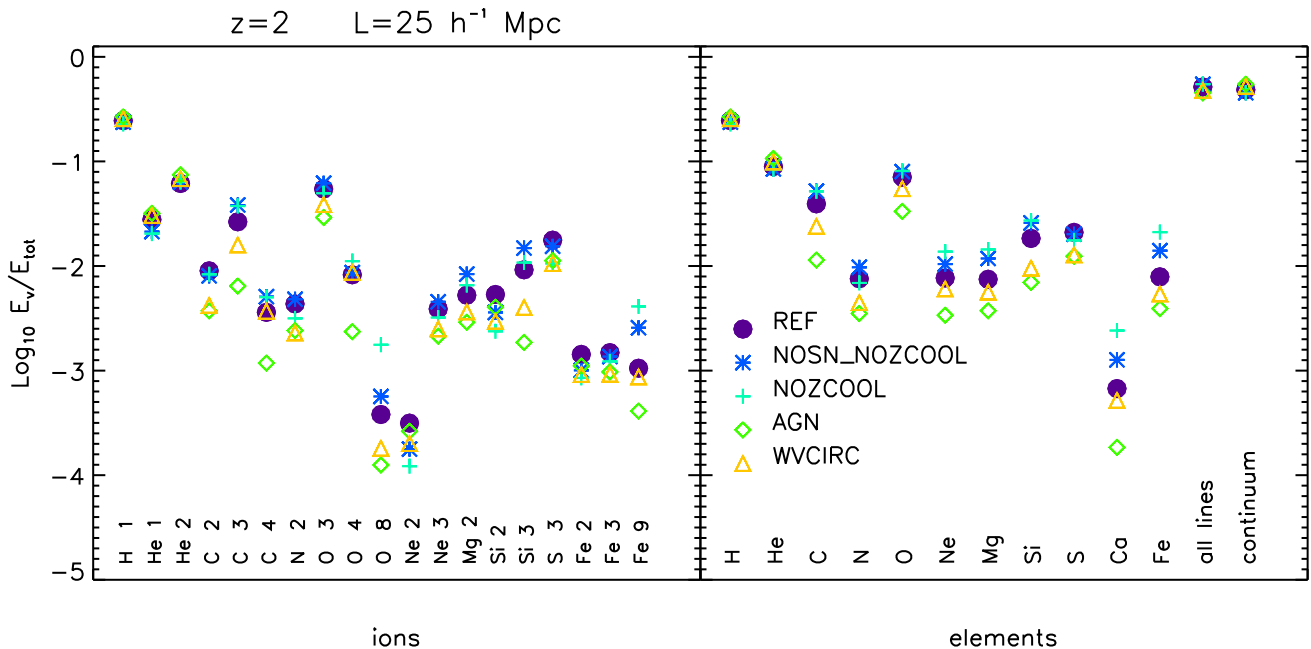}
\caption{As Fig.~\ref{sims_100}, but for a subset of \owls\ runs with boxes of 25 \hm\ Mpc at $z=2$.}
\label{sims_25}
\end{figure*}

In this Section we compare predictions from a subset of simulations from the \owls\ project with different implementations of a number of physical processes.
The full set of \owls\ simulations is described in detail in \citet{schaye2010} and we refer the interested reader to that work.

We use two subsets of simulations: a set with box sizes 100 \hm\ Mpc for results at $z=0$, and a set with box sizes 25 \hm\ Mpc for results at $z=2$. All runs contain $2\times 512^3$ particles and use the same particle masses and initial conditions as the reference runs. Wherever possible, we use simulations with 25 and 100 \hm\ Mpc boxes that implement the same physical prescriptions at both redshifts. In the following we only give a brief description of the runs we consider for our comparison.

\begin{enumerate}
\item \default: Our reference model, as described in Section~\ref{owlproj}.

\item \zcool: As the \default\ model, but with gas cooling rates calculated assuming primordial element abundances. The neglect of the contributions of metal lines implies slower gas cooling than in the \default\ model. Note that we do not turn off metal-line cooling in post processing the simulation to predict the emission, which is thus not self-consistent.

\item \nosn: As the \default\ model, but without supernova feedback: the metals produced by stars are only transported by gas mixing and no energy is transferred to the IGM and ISM when SNe explode (100 \hm\ Mpc box only).

\item \nosnz: As the \default\ model, but without supernova feedback and with gas cooling rates for primordial element abundances (25 \hm\ Mpc box only).

\item \noagb: As the \default\ model, but with no chemical enrichment by supernova Ia and AGB stars and no supernova Ia energy transfer (100 \hm\ Mpc box only).

\item \agn: As the \default\ model, with the addition of AGN feedback implemented according to the prescriptions of \citet{booth2009}.

\item \wmom: Model with a momentum--driven implementation of galactic winds, following the prescriptions of \citet{oppenheimer2008}, with the difference that winds are produced ``locally'' to the star formation event and are fully hydrodynamically coupled as in the \default\ model.
The wind initial velocity and the wind mass loading factor are defined as $v_{\rm w} = 3\sigma$ and $\eta = v_{\textrm{w}0}/\sigma$ respectively, with $\sigma = \sqrt{2}v_{\rm circ}$ the galaxy velocity dispersion, $v_{\rm circ}$ the halo circular velocity, estimated using an on-the-fly friends-of-friends halo finder, and $v_{\textrm{w}0}=150$ \kms.
\end{enumerate}

Fig.~\ref{sims_100} shows the specific emission of ions, elements and the continuum for simulations with box sizes 100 \hm\ Mpc at $z=0$.
Fig.~\ref{sims_25} shows the same quantities for simulations with box sizes 25 \hm\ Mpc at $z=2$.
The left panels show the specific emission of the ions most relevant to the global cooling rate, while the right panel shows the contributions of individual elements, all lines combined and the continuum.

Qualitatively, Figs.~\ref{sims_100} and \ref{sims_25} confirm the results of Fig.~\ref{fracions_100} that lines carry a larger fraction of the energy radiated away by cooling processes than the continuum, and that most of the cooling is produced by hydrogen lines. Variations in the relative contributions of lines and the continuum to the global budget are minimal for all models, with the expception of the \nosn\ model. Values of the specific emission of ions and elements vary among simulations by up to two orders of magnitude in the most extreme cases, with the largest variations for the \zcool\ model at $z=0$ and for highly ionised species (e.g.\ \oviii).

The \zcool\ model implements a less efficient cooling prescription than other models. As a consequence, the gas temperature is artificially high, with the largest variations at low redshift, where the gas is on average more metal-rich and metal-line cooling is more severely underestimated.
While the relative fractions of energy in lines and the continuum do not show significant variations, the fraction of energy emitted by hydrogen and helium lines is higher in the \zcool\ model than in \default, while the fractions emitted by metals are lower by up to a dex. When ions are considered, the difference tends to be larger for highly ionised species that emit energy in the \xray\ and FUV bands than for the lowest ionisation species.

Results for the \nosnz\ model are roughly similar to those of \default\ at $z=2$. However, results for the \nosn\ model are significantly different from those of \default\ at $z=0$. Line emission has an overall stronger impact than continuum emission for this model, and the individual contributions to the budget of ions and elements present significant variations. In particular, some cooling channels (e.g.\ silicon lines) are enhanced, while others (e.g.\ oxygen and magnesium lines) are suppressed.

The contribution of metal-line emission to the global budget is noticeably lower for the \agn\ model than for \default. This is because AGN feedback quenches star formation and suppresses the production of metals, which translates into a lower specific emission for metal lines.

Variations in the emission channels due to different feedback mechanisms are smaller than for other physical variations. This is demonstrated by the results for the \wmom\ model, which are on average within a factor of a few of those of \default\ and present the smallest variations in our sample of models.
This means that the way metals are injected in to the IGM through supernova feedback is largely irrelevant in the determination of the global cooling channels.

In the \noagb\ model, the production of metals by AGB stars and type Ia supernovae is turned off, which implies that the relative abundances of some elements, such as carbon and iron, are lower than in the \default\ model. This directly translates in to smaller relative contributions to the gas cooling by these same elements, with differences of up to 0.5 dex.

In conclusion, the comparison between different simulations demonstrates that: i) The gas cooling rates are the most essential ingredient to determine the gas temperature and ultimately the relative importance of different cooling channels. Since the cooling rates depend on metallicity, which on average increases with decreasing redshift, the underestimation of the gas cooling rate when assuming primordial abundances becomes more severe with decreasing redshift. ii) The largest differences between the predictions of different simulations are visible for highly ionised species, such as \oviii. This confirms the findings of \citet{bertone2010a} that \xray\ observations might provide a strong tool to rule out or verify the theoretical predictions from different models.

\section{Conclusions}
\label{conclusion}

In this work we have investigated the cooling channels of diffuse gas in the Universe. We limited our analysis to cooling radiation from diffuse, non star forming gas with $n_{\rm H} < 0.1$ cm\3, and we neglected the emission from stars, supernova explosions, active galactic nuclei and compact objects.

We used a subset of high resolution, cosmological hydrodynamical simulations from the \owls\ project \citep{schaye2010}, with box sizes of 25 \hm\ Mpc and 100 \hm\ Mpc on the side and $2\times 512^3$ particles. Lower resolution simulations have been used to test the dependence of the results on resolution and box size (see Appendix).
Most of our results were inferred rom the \default\ simulations, which include element-by-element cooling for 11 species, star formation, stellar evolution and chemodynamics, and galactic winds driven by star formation.

Our main results can be summarised as follows:

\begin{enumerate}
\item Emission lines carry about 80 per cent (70 per cent) of the energy emitted by cooling radiation at redshift $z=2$ ($z=0$), with the remainder carried by continuum emission. 
Of the energy emitted by lines, the majority is carried by \lya\ and other Lyman series lines of hydrogen. Other important coolants are oxygen, helium and carbon at $z=2$, and silicon, carbon, iron and helium at $z=0$, in order of decreasing contributions. This shift in the relative contributions of different elements reflects variations in the production of metals with time: while oxygen from  massive stars is the most abundant metal at high redshift, the production of carbon and iron by AGB stars and supernova Ia explosions is boosted at lower redshifts \citep{wiersma2009b}.
In addition, structure formation increases the amount of hot gas with time, favouring the line emission from elements with higher atomic numbers, such as iron.

\item The emission lines that carry most of the energy usually correspond to relatively low ionisation states of atoms. In particular, some of the strongest lines correspond to \oiii, \cii, \ciii, \silii, \siliii, \feii\ and \siii\ ions. The wavelengths of the strongest lines of each ion can be read in Figs.~\ref{emissivity1} and \ref{emissivity2} and the strongest lines for each element are shown in Figs. \ref{lines_vs_temp_z=0} and \ref{lines_vs_temp_z=2} and listed in Appendix \ref{linelist}.

\item Emission in the UV band strongly dominates the energy budget at all redshifts. With almost no exceptions, the strongest emission lines are UV lines. Overall, the contribution of the continuum emission to the energy budget is larger than the contribution of line emission at high energies (\xray\ and FUV bands), while lines dominate at lower energies (NUV, optical and IR+ bands).
\xray\ emission carries a few per cent of the total energy, its contribution increasing with decreasing redshift. The optical and IR+ bands together contribute less than 20 per cent of the total emission, with IR+ lines dominating the budget at $z=0$.

\item We have compared results for a number of \owls\ runs that implement variations of a selected number of physical prescriptions.
We find that the gas cooling rates are the most essential ingredient to determine the gas temperature and ultimately the relative importance of different cooling channels. Since the cooling rates depend on the gas metallicity, which on average increases with decreasing redshift, the underestimation of the gas cooling rate when assuming primordial abundances becomes more severe at low redshift. The largest variations in the predictions preferentially affect highly ionised species, such as \oviii.

\end{enumerate}

Our results demonstrate that the UV band is a treasure trove of information about the cooling processes that determine the evolution of the diffuse IGM. Current IGM studies mostly target absorption lines in QSO spectra such as \lya, \civ, \ovi. These lines are extremely useful and are relatively easy to detect. However, \civ\ and \ovi\ lines only carry a small fraction of the total energy radiated away by cooling gas, compared with other UV lines from lower ionisation states, such as \ciii, \oiii\ and \ov, which account for a much larger portion of the energy budget.

The fact that most of the cooling radiation is emitted in the UV band, both by the continuum and by emission lines, may have some interesting consequences. In fact, photons emitted at wavelengths shorter than the Lyman limit break at 912 \AA\ have a very short mean free path at high redshift (e.g.\ \citealt{inoue2008}; \citealt{prochaska2009}). As such, a large fraction of the energy originally emitted in the FUV band is quickly absorbed and re-radiated at longer wavelengths and it might be problematic to detect cooling radiation from diffuse gas directly.

At low redshift, however, the mean free path of ionising photons increases substantially, and the Universe becomes more transparent (e.g.\ \citealt{madau1999}). This means that a larger fraction of the cooling radiation in the FUV band could be detected. Indeed, evidence of FUV emission in clusters has been detected by EUVE observations \citep{lieu1996}.
It would therefore be highly desirable to build a satellite with the capability to try to detect emission at high spectral resolution in the FUV band $\lambda \sim 100-1000$ \AA, where most of the strongest emission lines and a large fraction of the continuum emission are found.

\section*{Acknowledgments}
We would like to thank the Referee for the timely and kind report.
The simulations presented here were run on Stella, the LOFAR BlueGene/L system in Groningen, on the Cosmology Machine at the Institute for Computational Cosmology
in Durham (which is part of the DiRAC Facility jointly funded by STFC, the Large Facilities Capital Fund of BIS, and Durham University) as part of the Virgo Consortium research programme, and on Darwin in Cambridge. This work was sponsored by the National
Computing Facilities Foundation (NCF) for the use of supercomputer facilities, with financial support from the Netherlands Organization for Scientific Research (NWO), also through a VIDI grant.
SB and AA acknowledge support by NSF Grants AST-0507117 and AST-0908910.
JS acknowledge support from the European Research Council under the European Union’s Seventh Framework Programme (FP7/2007-2013) / ERC Grant agreement 278594-GasAroundGalaxies and from the Marie Curie Training Network CosmoComp (PITN-GA-2009-238356).

\appendix

\section{Convergence tests}
\label{converge}

In this Appendix we investigate how resolution (Appendix \ref{number}) and box size (Appendix \ref{boxsize}) affect our results.
We show results for the specific emission of oxygen ions, which are broadly representative of all elements in all bands. Equivalent results hold for all elements traced by \owls.
In Appendix \ref{nbins} we show how the binning in wavelength used to calculate the continuum emission affects the estimate of the continuum contribution to the total energy budget.
All results are shown for runs that use the \default\ physical model.

\subsection{Resolution}
\label{number}

\begin{figure*}
\centering
\includegraphics[width=8.4cm]{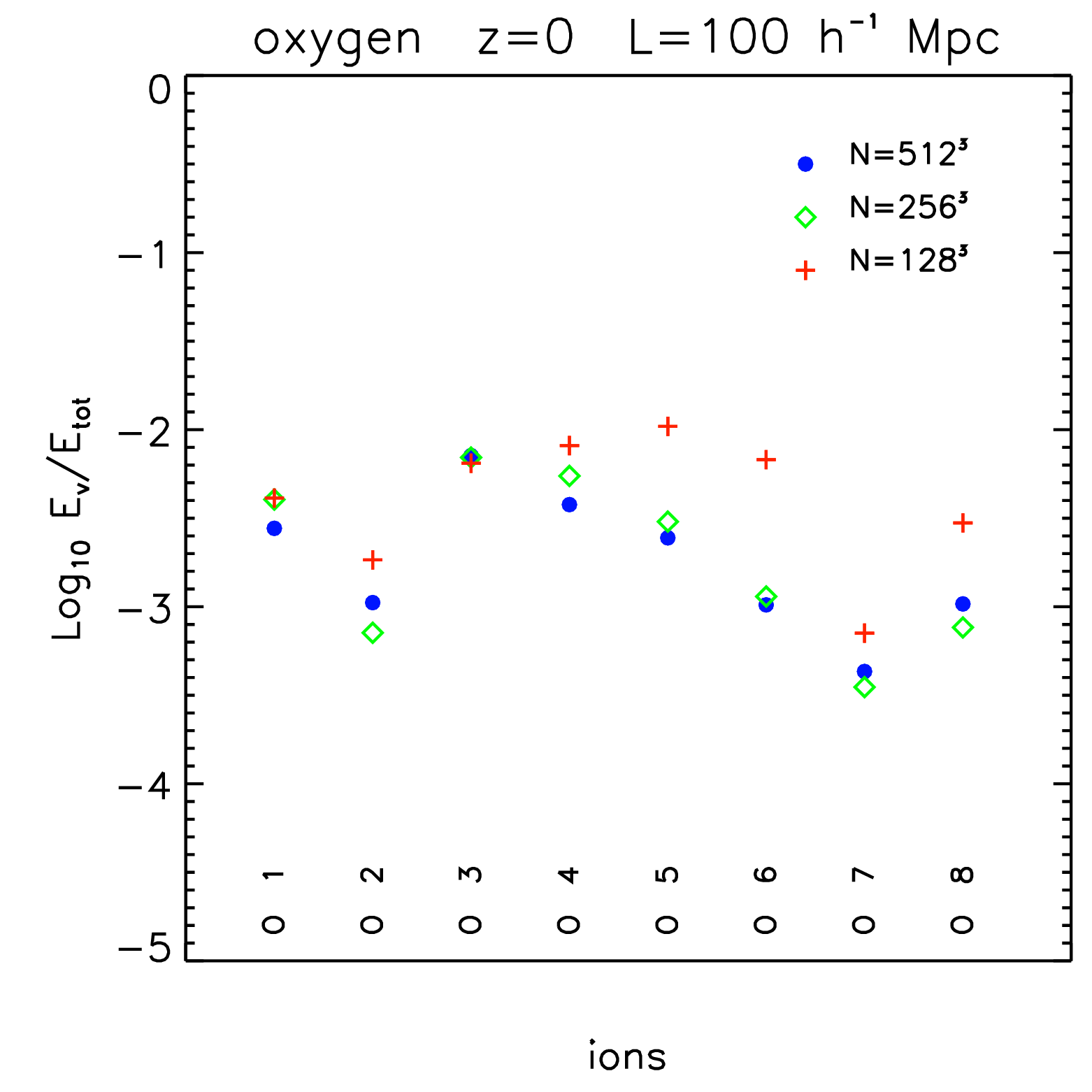}
\includegraphics[width=8.4cm]{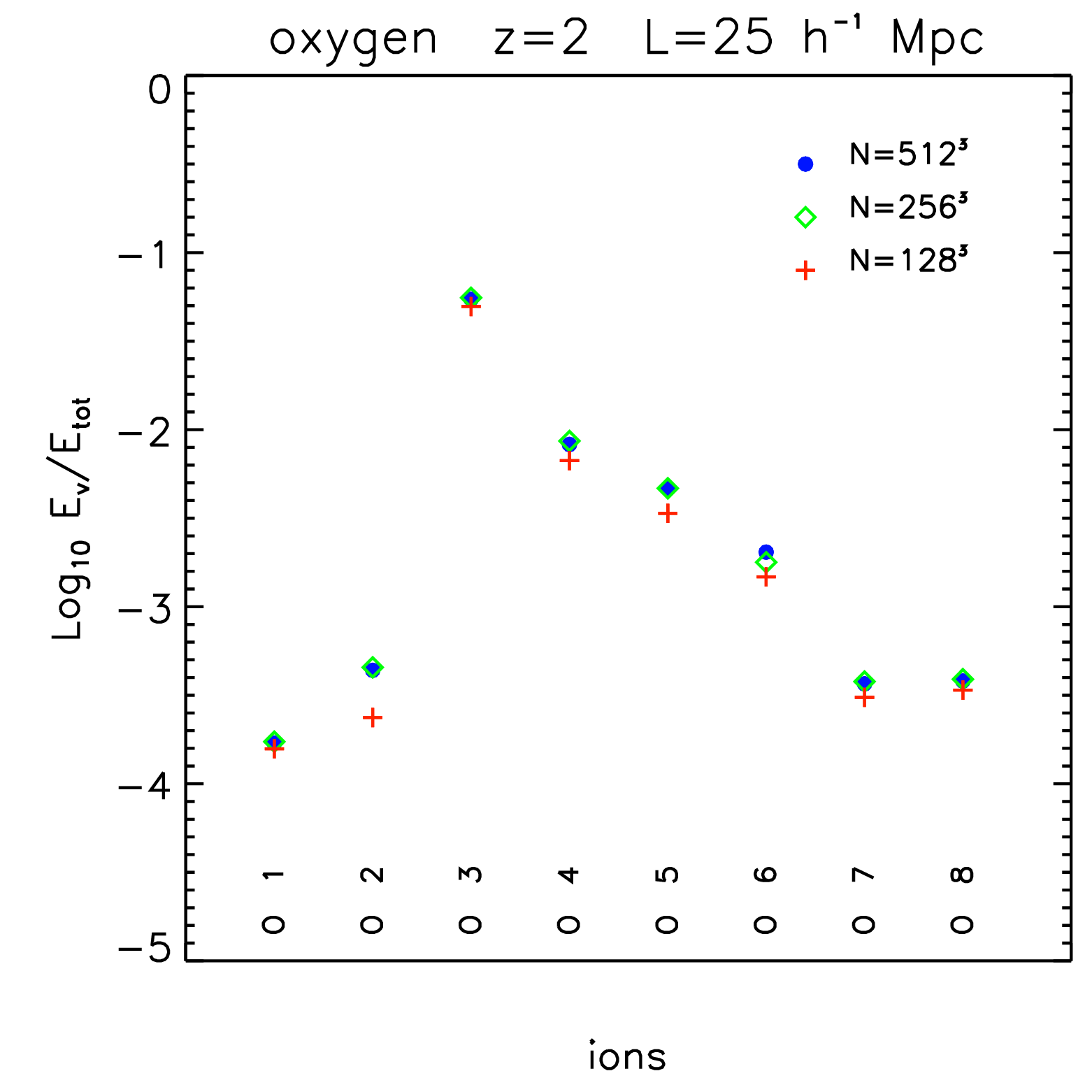}
\caption{Dependence of the specific emission (emission per unit volume) on mass resolution. Results are shown for runs with $L=25$ \hm\ Mpc at $z=2$ (left panel) and for runs with $L=100$ \hm\ Mpc at $z=0$ (right panel), for constant box size. The number of particles in the simulations is $N=2\times 128^3$, $2\times 256^3$ and $2\times 512^3$, corresponding to DM particles masses $M_{\rm DM} = 4.1\times 10^{8}$ \hm\ \msun, $M_{\rm DM} = 5.1\times 10^7$ \hm\ \msun\ and $M_{\rm DM} = 6.3\times 10^6$ \hm\ \msun\  (left panel) and to $M_{\rm DM} = 2.6\times 10^{10}$ \hm\ \msun, $M_{\rm DM} = 3.2\times 10^9$ \hm\ \msun\ and $M_{\rm DM} = 4.1\times 10^8$ \hm\ \msun\ (right panel).}
\label{massz0}
\end{figure*}

We investigate the effect of varying the numerical resolution by comparing results for the specific emission of oxygen ions in simulations with varying particle number and constant box sizes. The simulations contain $N=2\times 128^3$, $2\times 256^3$ and $2\times 512^3$ particles each. As a consequence, the mass and the spatial resolution vary in steps of 8 and 2, respectively

Fig.~\ref{massz0} shows results for simulations with 25 \hm\ Mpc boxes and DM particle masses $M_{\rm DM} = 4.1\times 10^{8}$ \hm\ \msun, $5.1\times 10^7$ \hm\ \msun\ and $6.3\times 10^6$ \hm\ \msun\ at $z=2$ (left panel) and for simulations with 100 \hm\ Mpc boxes and DM particle masses $M_{\rm DM} = 2.6\times 10^{10}$ \hm\ \msun, $3.2\times 10^9$ \hm\ \msun\ and $4.1\times 10^8$ \hm\ \msun\ at $z=0$ (right panel).

Results are well converged at $z=2$ for all particle masses and deviations among different sets of simulations are minimal. At $z=0$ only the two higher resolution runs converge well, while the run with $N=2\times 128^3$ particles diverges significantly for most ions. This is not surprising, since neither the star formation rate density \citep{schaye2010} nor the line emission (e.g.\ \citealt{bertone2010a}) converge for this simulation.
The convergence in the 25 \hm\ Mpc simulations at $z=2$ for particle masses equal or smaller than those used in the 100 \hm\ Mpc run with $N=2\times 512^3$ particles, suggests that a good level of convergence is reached in the highest resolution run with a box size of 100 \hm\ Mpc.

\subsection{Box size}
\label{boxsize}

\begin{figure}
\centering
\includegraphics[width=8.4cm]{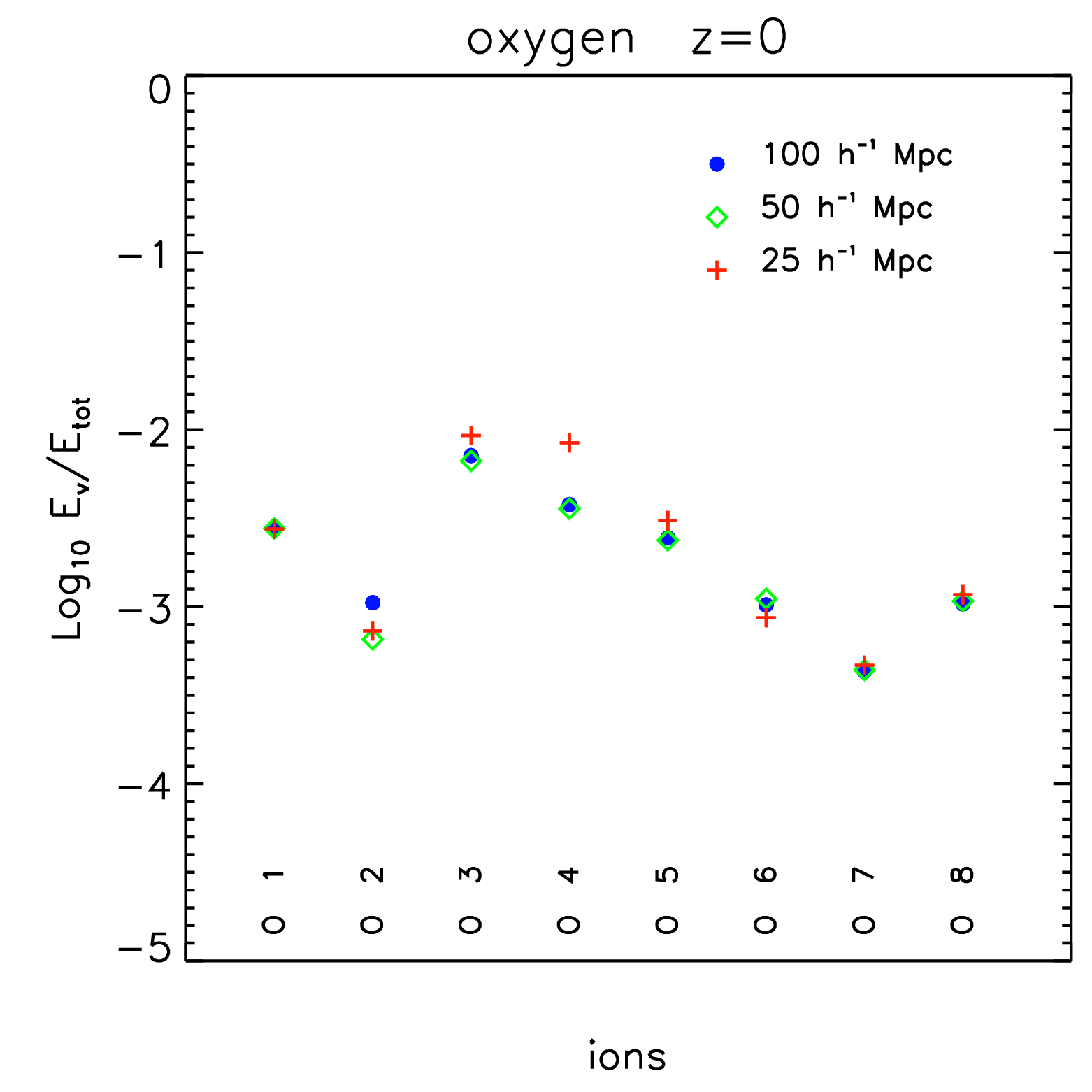}
\caption{Dependence of results on the box dimension $L$. Results are shown at $z=0$ for simulations with constant resolution and box sizes of 25, 50 and 100 \hm\ Mpc.}
\label{box}
\end{figure}

In this Appendix we investigate the effect of varying the box dimension at constant mass resolution.
We compare results for simulations that have DM particle masses $M_{\rm DM} = 4.1\times 10^8$ \hm\ \msun\ and box sizes of 25, 50 and 100 \hm\ Mpc at $z=0$. The number of particles in each run is $N=2\times 128^3$, $N=2\times 256^3$ and $N=2\times 512^3$, respectively.

Results for the specific emission of oxygen are shown in Fig.~\ref{box} and show good convergence for all box sizes.
\oiv\ emission shows variations of about 0.5 dex for the 25 \hm\ Mpc run, also found in nitrogen and carbon ions at equivalent ionisation temperatures.
Small differences are seen in highly ionised species of calcium and iron (e.g.\ \fexxv\ and \fexxvi), which emit in the hard \xray\ band. These are due to the relatively small box size of the 25 \hm\ Mpc run, which does not contain the massive groups and clusters in which gas can reach temperatures high enough to produce the highest ionisation states of atoms with high atomic numbers.
Variations for other ions are within 0.2 dex or less.

\subsection{Resolution in the continuum spectrum}
\label{nbins}

\begin{figure}
\centering
\includegraphics[width=8.4cm]{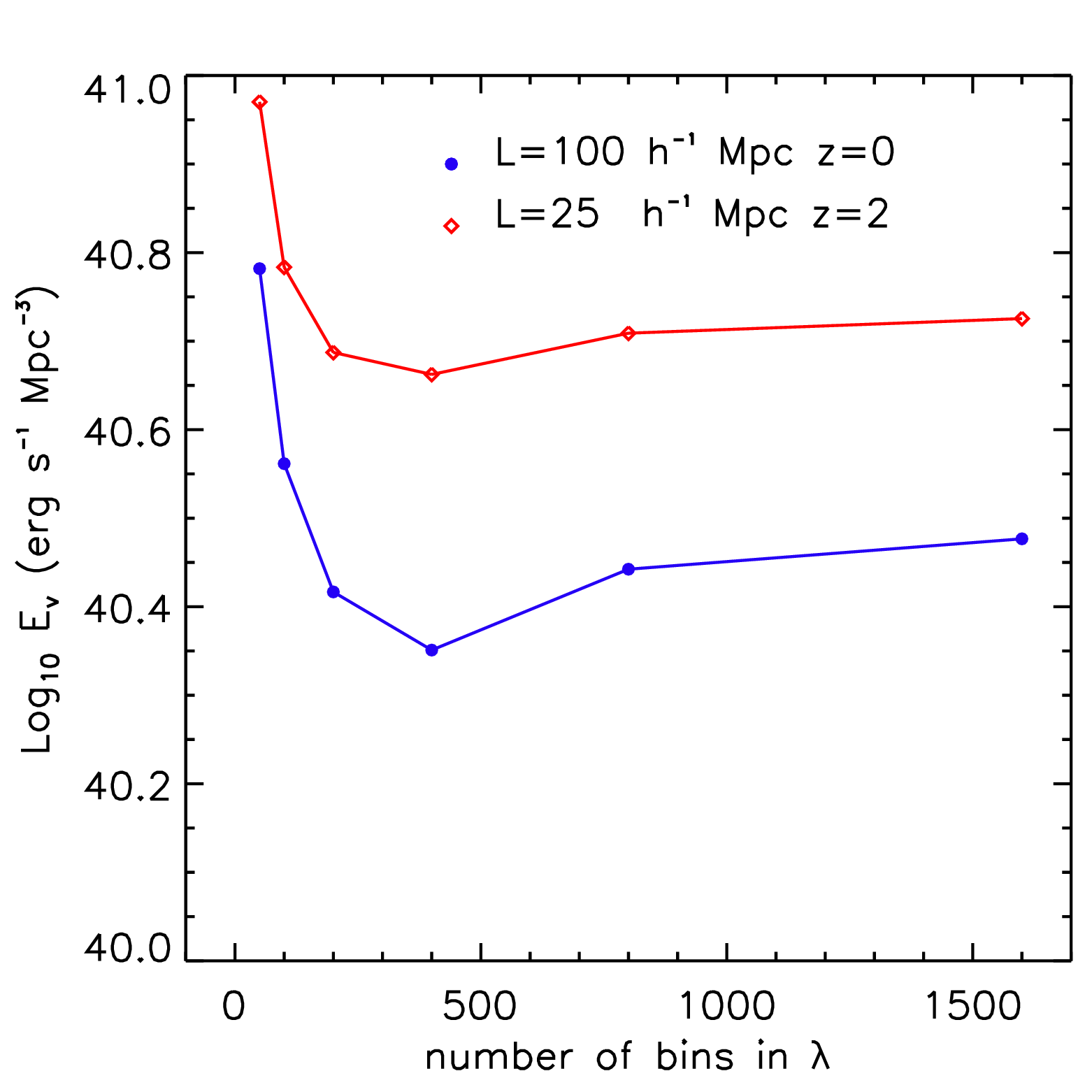}
\caption{Dependence of the specific emission (emission per unit volume) of the continuum on the number of bins used to integrate the intensity of the continuum. Differences in the results are well within 0.4 dex.}
\label{nlambda}
\end{figure}

In this Appendix we investigate how the resolution in wavelength used to interpolate the continuum tables affects the estimate of the contribution of the continuum to the total energy budget.

\cloudy\ provides the intensity of the continuum emissivity as a function of wavelength in the range $\lambda =$0.01 \AA -1 m. The binning in wavelength is irregular, with smaller bins at shorter wavelengths. At full resolution, the continuum emissivity contains $N_{\rm bin}\sim 3700$ elements. As shown in Fig.~\ref{contemp}, the spectral shape of the continuum emissivity contains sharp edges at low temperatures. These need to be resolved when interpolating the continuum to create full emission spectra. However, for our purposes this might not be necessary, since we are only interested in estimating the total energy carried by the continuum.
To speed up our calculations, we have binned the continuum in a smaller number of wavelengths than those contained in the tables, and we perform the interpolation over this new set of wavelengths.

To verify that the binning does not affect the estimate of the specific emission of the continuum, we have tested how this varies as a function of $N_{\rm bin}$. Results for runs with 25 \hm\ Mpc at $z=2$ and with 100 \hm\ Mpc at $z=0$ are shown in Fig.~\ref{nlambda} for $N_{\rm bin} =$50, 100, 200, 400, 800 and 1600. We find that a high number of bins does not substantially improve the accuracy in the estimate of the specific emission. Variations in the results are within 0.3 dex, and decrease to within 0.1 dex or less for $N_{\rm bin}>50$.
For this work, we assume $N_{\rm bin} = 200$ everywhere, except for Fig.~\ref{redshift}, where we use $N_{\rm bin} = 100$.

\section{Line lists}
\label{linelist}

\begin{table*}
\centering
\caption{List of the individual emission lines that account for most of the total emitted energy. The columns contain:
i) the ionic species;
ii) the wavelength of the line in \AA;
iii) the temperature $T_{\rm m}$ at which the line carries the largest fraction of the total energy;
iv) and viii) the total fraction $F$ of energy radiated through the line at $z=0$ and $z=2$;
v) and ix) the rank of the line according to the total fraction of radiated energy.
vi) and x) the fraction $F_{\rm m}$ of the energy carried by the line at the temperature $T_{\rm m}$ where it carries the largest fraction of the total energy,  at $z=0$ and $z=2$;
vii) and xi) the rank of the line according to the fraction of radiated energy at $T_{\rm m}$.
}
\begin{tabular}{l r c r r r r r r r r}
\hline
& & & \multicolumn{4}{c}{$z=0$} & \multicolumn{4}{c}{$z=2$}\\
Line & $\lambda$ (\AA) & Log $T_{\rm m}$ (K) & $F$ & Rank & $F(T_{\rm m})$ & Rank & $F$ & Rank & $F(T_{\rm m})$ & Rank \\
\hline
\hi\      & 1216   & 4.2 & 11.26 & 1   & 26.1 & 2  & 12.96 & 1  & 17.6 & 3 \\
\hi\      & 1026   & 3.0 & 5.35  & 2   & 9.1  & 12 & 4.27  & 2  & 6.9  & 12 \\
\hi\      & 972.5  & 3.0 & 2.86  & 4   & 5.6  & 23 & 2.00  & 5  & 4.3  & 20 \\
\hi\      & 949.7  & 3.0 & 1.77  & 6   & 3.9  & 33 & 1.15  & 6  & 2.9  & 32 \\
\hi\      & 937.8  & 3.0 & 1.18  & 7   & 2.8  & 50 & 0.69  & 10 & 2.1  & 43 \\
\hi\      & 930.8  & 3.0 & 0.84  & 8   & 2.1  & 60 & 0.47  & 13 & 1.6  & 54 \\
\hi\      & 926.2  & 3.0 & 0.62  & 11  & 1.6  & 73 & 0.34  & 19 & 1.2  & 62 \\
\hi\      & 923.2  & 3.0 & 0.48  & 14  & 1.3  & 84 & --    & -- & --   & -- \\
\hi\      & 921.0  & 3.0 & 0.38  & 16  & 1.1  & 90 & --    & -- & --   & -- \\
\hi\      & 6563   & 3.0 & 0.66  & 9   & 1.1  & 88 & --    & -- & --   & -- \\

\hei\     & 625.6  & 3.0 & 2.16  & 5   & 3.5  & 38 & 2.00  & 4  & 6.4  & 13\\
\hei\     & 584.3  & 3.0 & --    & --  & --   & -- & 0.38  & 17 & 1.4  & 59\\
\heii\    & 303.8  & 4.9 & 0.32  & 18  & 10.1 & 10 & 2.88  & 3  & 24.2 & 2 \\
\heii\    & 256.3  & 5.1 & 0.08  & 40  & 1.1  & 87 & 1.11  & 8  & 3.0  & 31\\

\cii\     & 1335   & 4.5 & 0.27  & 21  & 6.5  & 17 & 0.20  & 26 & 1.0  & 70 \\
\cii\     & 892.8  & 4.5 & 0.06  & 43  & 2.0  & 62 & --    & -- & --   & -- \\
\cii\     & 2328   & 4.4 & 0.21  & 29  & 2.4  & 53 & --    & -- & --   & -- \\
\ciii\    & 977.0  & 4.8 & 0.64  & 10  & 25.5 & 3  & 1.12  & 7  & 9.9  & 8  \\
\ciii\    & 1907   & 4.6 & 0.26  & 24  & 5.4  & 25 & 0.91  & 9  & 2.5  & 37 \\
\ciii\    & 1910   & 4.6 & 0.17  & 30  & 3.5  & 37 & 0.60  & 11 & 1.6  & 51 \\
\civ\     & 1548   & 5.0 & 0.09  & 39  & 8.1  & 13 & 0.24  & 24 & 3.9  & 23 \\
\civ\     & 1551   & 5.0 & 0.04  & 52  & 4.1  & 31 & 0.12  & 34 & 1.9  & 48 \\

\niii\    & 991.0  & 4.8 & 0.05  & 48  & 1.8  & 65 & --    & -- & --   & -- \\
\niii\    & 678.2  & 4.9 & 0.03  & 64  & 1.4  & 80 & --    & -- & --   & -- \\
\niv\     & 765.0  & 5.1 & 0.06  & 42  & 7.1  & 16 & 0.05  & 45 & 2.5  & 38 \\
\nv\      & 1239   & 5.2 & 0.01  & 86  & 1.6  & 75 & --    & -- & --   & -- \\

\oii\     & 833.8  & 4.6 & 0.10  & 35  & 3.7  & 36 & --    & -- & --   & -- \\
\oiii\    & 698.2  & 4.9 & 0.12  & 33  & 6.4  & 18 & 0.20  & 25 & 3.2  & 27 \\
\oiii\    & 500.7  & 5.0 & 0.04  & 56  & 2.5  & 52 & 0.05  & 46 & 1.3  & 61 \\
\oiii\    & 835.0  & 4.9 & 0.04  & 55  & 1.8  & 67 & --    & -- & --   & -- \\
\oiv\     & 549.3  & 5.2 & 0.14  & 31  & 15.9 & 5  & 0.25  & 23 & 11.8 & 4  \\
\oiv\     & 789.0  & 5.2 & 0.13  & 32  & 12.3 & 9  & 0.26  & 21 & 9.3  & 10 \\
\oiv\     & 609.4  & 5.2 & 0.04  & 50  & 4.8  & 27 & 0.08  & 39 & 3.5  & 26 \\
\ov\      & 630.0  & 5.4 & 0.23  & 27  & 37.4 & 1  & 0.43  & 14 & 30.9 & 1  \\
\ov\      & 1218   & 5.3 & 0.01  & 89  & 1.3  & 83 & 0.02  & 69 & 1.1  & 67 \\
\ovi\     & 1032   & 5.5 & 0.07  & 41  & 12.4 & 8  & 0.13  & 32 & 9.9  & 9  \\
\ovi\     & 1038   & 5.5 & 0.03  & 58  & 6.2  & 21 & 0.07  & 43 & 5.0  & 17 \\
\ovii\    & 21.60  & 6.3 & 0.02  & 77  & 1.8  & 66 & 0.01  & 82 & 2.1  & 44 \\
\ovii\    & 22.10  & 6.3 & 0.02  & 71  & 1.9  & 64 & 0.02  & 77 & 2.2  & 40 \\
\oviii\   & 18.97  & 6.5 & 0.09  & 38  & 3.8  & 34 & 0.03  & 57 & 5.2  & 16 \\

\neiv\    & 546.2  & 5.2 & 0.02  & 72  & 2.0  & 61 & 0.03  & 62 & 1.2  & 65 \\
\nev\     & 573.4  & 5.5 & 0.02  & 73  & 3.2  & 42 & 0.03  & 59 & 2.0  & 46 \\
\nev\     & 481.2  & 5.5 & 0.01  & 81  & 2.7  & 51 & 0.02  & 71 & 1.7  & 49 \\
\nevi\    & 400.0  & 5.6 & 0.05  & 47  & 14.0 & 7  & 0.07  & 40 & 10.0 & 7  \\
\nevi\    & 561.8  & 5.6 & 0.01  & 88  & 3.0  & 45 & 0.02  & 78 & 2.1  & 42 \\
\nevi\    & 432.0  & 5.6 & 0.01  & 100 & 2.0  & 63 & 0.01  & 91 & 1.4  & 58 \\
\nevii\   & 464.9  & 5.7 & 0.03  & 59  & 9.0  & 11 & 0.05  & 48 & 7.5  & 11 \\
\neviii\  & 770.4  & 5.8 & 0.02  & 78  & 3.1  & 44 & 0.02  & 68 & 2.7  & 35 \\
\neviii\  & 780.3  & 5.8 & 0.01  & 93  & 1.5  & 76 & 0.01  & 85 & 1.3  & 60 \\

\mgii\    & 2796   & 4.1 & 0.53  & 12  & 4.3  & 30 & --    & -- & --   & -- \\
\mgii\    & 2803   & 4.1 & 0.27  & 22  & 2.1  & 58 & --    & -- & --   & -- \\
\mgvi\    & 406.4  & 5.6 & 0.01  & 85  & 3.3  & 41 & 0.03  & 61 & 3.1  & 28 \\
\mgvii\   & 276.9  & 5.8 & 0.01  & 96  & 1.8  & 68 & 0.02  & 79 & 2.0  & 47 \\
\mgviii\  & 315.5  & 5.9 & 0.02  & 70  & 3.9  & 32 & 0.03  & 55 & 4.4  & 19 \\
\mgix\    & 368.5  & 5.9 & 0.01  & 82  & 2.3  & 55 & 0.02  & 72 & 2.7  & 34 \\
\hline
\end{tabular}
\label{linelist1}
\end{table*}

\begin{table*}
\centering
\caption{As Fig. \ref{linelist1} - continued}
\begin{tabular}{l r c r r r r r r r r}
\hline
& & & \multicolumn{4}{c}{$z=0$} & \multicolumn{4}{c}{$z=2$}\\
Line & $\lambda$ (\AA) & Log $T_{\rm m}$ (K) & $F$ & Rank & $F(T_{\rm m})$ & Rank & $F$ & Rank & $F(T_{\rm m})$ & Rank \\
\hline

\silii\   & 348100 & 3.0 & 4.29  & 3   & 16.2 & 4   & 0.40 & 16 & 11.7 & 5  \\
\siliii\  & 1207   & 4.5 & 0.26  & 25  & 7.5  & 15  & 0.27 & 20 & 1.7  & 50 \\
\siliii\  & 1883   & 4.4 & 0.28  & 20  & 5.6  & 22  & 0.41 & 15 & 1.2  & 64 \\
\siliv\   & 1394   & 4.8 & 0.06  & 44  & 2.8  & 49  & --   & -- & --   & -- \\
\siliv\   & 1403   & 4.8 & 0.03  & 61  & 1.4  & 81  & --   & -- & --   & -- \\
\silvi\   & 249.1  & 5.6 & 0.01  & 105 & 1.2  & 85  & --   & -- & --   & -- \\
\silvii\  & 277.6  & 5.8 & 0.02  & 75  & 3.4  & 40  & 0.02 & 64 & 2.7  & 33 \\
\silviii\ & 319.1  & 5.9 & 0.04  & 53  & 6.4  & 19  & 0.05 & 51 & 5.2  & 15 \\
\silix\   & 354.2  & 6.0 & 0.02  & 68  & 3.0  & 46  & 0.02 & 70 & 2.6  & 36 \\
\silx\    & 260.6  & 6.1 & 0.02  & 66  & 3.5  & 39  & 0.02 & 73 & 3.1  & 29 \\
\silxi\   & 303.9  & 6.2 & 0.03  & 60  & 4.5  & 28  & 0.03 & 63 & 4.2  & 21 \\
\silxii\  & 499.0  & 6.3 & 0.02  & 69  & 1.7  & 70  & 0.01 & 87 & 1.6  & 53 \\

\siii\    & 667.0  & 4.8 & 0.05  & 45  & 2.2  & 57  & --   & -- & --   & -- \\
\sv\      & 786.5  & 5.1 & 0.02  & 67  & 3.1  & 43  & 0.02 & 65 & 1.5  & 57 \\
\six\     & 228.8  & 6.1 & 0.01  & 80  & 1.7  & 72  & 0.01 & 80 & 1.5  & 56 \\
\sx\      & 265.9  & 6.2 & 0.02  & 74  & 2.3  & 56  & 0.02 & 76 & 2.2  & 41 \\
\sxii\    & 222.2  & 6.3 & 0.01  & 87  & 1.7  & 71  & 0.01 & 96 & 1.6  & 52 \\
\sxiii\   & 262.7  & 6.4 & 0.03  & 62  & 3.8  & 35  & 0.01 & 81 & 3.8  & 24 \\

\cavi\    & 330.4  & 5.6 & 0.01  & 98  & 1.8  & 69  & 0.01 & 90 & 1.2  & 63 \\
\cavii\   & 338.0  & 5.6 & 0.01  & 111 & 1.1  & 89  & --   & -- & --   & -- \\

\feii\    & 2400   & 4.1 & 0.33  & 17  & 2.4  & 54  & --   & -- & --   & -- \\
\feii\    & 259900 & 3.1 & 0.48  & 13  & 1.4  & 79  & --   & -- & --   & -- \\
\fevii\   & 163.5  & 5.6 & 0.01  & 101 & 1.4  & 82  & --   & -- & --   & -- \\
\feviii\  & 160.8  & 5.7 & 0.01  & 94  & 1.5  & 77  & 0.01 & 89 & 1.1  & 69 \\
\feix\    & 169.7  & 5.9 & 0.10  & 36  & 15.9 & 6   & 0.10 & 38 & 10.8 & 6  \\
\fex\     & 171.8  & 6.0 & 0.05  & 46  & 7.7  & 14  & 0.04 & 52 & 5.6  & 14 \\
\fex\     & 174.3  & 6.0 & 0.03  & 63  & 4.3  & 29  & 0.02 & 66 & 3.0  & 30 \\
\fex\     & 184.8  & 6.0 & 0.01  & 90  & 1.6  & 74  & 0.01 & 93 & 1.1  & 66 \\
\fexi\    & 178.5  & 6.1 & 0.03  & 57  & 5.5  & 24  & 0.03 & 60 & 4.0  & 22 \\
\fexi\    & 188.2  & 6.1 & 0.01  & 92  & 1.5  & 78  & 0.01 & 97 & 1.1  & 68 \\
\fexii\   & 193.4  & 6.1 & 0.04  & 51  & 6.4  & 20  & 0.03 & 58 & 4.7  & 18 \\
\fexiii\  & 204.3  & 6.2 & 0.02  & 76  & 3.0  & 47  & 0.01 & 86 & 2.2  & 39 \\
\fexiv\   & 219.8  & 6.3 & 0.01  & 83  & 2.1  & 59  & 0.01 & 95 & 1.5  & 55 \\
\fexv\    & 284.2  & 6.3 & 0.05  & 49  & 5.0  & 26  & 0.02 & 75 & 3.5  & 25 \\
\fexvi\   & 335.4  & 6.5 & 0.04  & 54  & 2.8  & 48  & 0.01 & 88 & 2.0  & 45 \\
\fexx\    & 121.0  & 7.0 & 0.02  & 65  & 1.1  & 86  & --   & -- & --   & -- \\
\hline
\end{tabular}
\label{linelist2}
\end{table*}

In this Appendix we provide the full lists of the individual emission lines that contribute most of the cooling energy. The fractions of the mean energy emitted per unit volume for individual lines as a function of temperature are shown in Figs. \ref{lines_vs_temp_z=0} and \ref{lines_vs_temp_z=2}. Only lines that contribute at least 1 per cent of the total emitted energy in at least one temperature bin are listed. Table \ref{linelist1} lists line for H, He, C, N, O, Ne and Mg. Table \ref{linelist2} lists lines for Si, S, Ca and Fe.

\bsp
\label{lastpage}

\end{document}